\documentstyle[a4,12pt,epsf]{article}
\epsfverbosetrue

\newcommand{\be}{\begin{equation}}
\newcommand{\ee}{\end{equation}}
\newcommand{\bea}{\begin{eqnarray}}
\newcommand{\eea}{\end{eqnarray}}
\newcommand{\brr}{\begin{array}}
\newcommand{\err}{\end{array}}
\newcommand{\bc}{\begin{center}}
\newcommand{\ec}{\end{center}}

\def\ltap{\;\raisebox{-.5ex}{\rlap{$\sim$}} \raisebox{.5ex}{$<$}\;}
\def\gtap{\;\raisebox{-.5ex}{\rlap{$\sim$}} \raisebox{.5ex}{$>$}\;}

\newcommand{\zl}{Z_L}
\newcommand{\mr}{M_\rho}
\newcommand{\fpi}{f_\pi}
\newcommand{\ia}{a^{-1}}
\newcommand{\fb}{f_B}
\newcommand{\fbs}{f_B^{stat}}

\begin{document}

\setcounter{page}{1}
\begin{flushright}
ROME preprint 94/1041 \\
October, 1994. \\
\end{flushright}

\vskip 10mm
\centerline{\Large{\bf{The continuum limit of $\fb$ from the lattice}}}
\centerline{\Large{\bf{in the static approximation.}}}

\vskip 10mm
\centerline{\bf{C.R. Allton}}

\vskip 10mm
\centerline{\em{INFN, Sezione di Roma and Dipartimento di Fisica,}}
\centerline{\em{Universita di Roma \lq La Sapienza\rq,
P.le A. Moro, I-00185 Rome, Italy.}}

\vskip 10mm
\begin{abstract}
We present an analysis of the continuum extrapolation
of $f_B$ in the static approximation from lattice data.
The method described here aims to uncover the systematic
effects which enter in this extrapolation and has
not been described before.
Our conclusions are that we see statistical evidence
for scaling of $f_B^{stat}$ for inverse lattice spacings $\gtap 2$ GeV
but not for $\ltap 2$ GeV.
We observe a lack of {\em asymptotic} scaling for a variety of
quantities, including $f_B^{stat}$, at all energy scales considered.
This can be associated with finite lattice spacing systematics.
Once these effects are taken into account, we obtain a value of
230(35) MeV for $f_B^{stat}$ in the continuum
where the error represents uncertainties due to both the
statistics and the continuum extrapolation.
In this method there is no error due to uncertainties in
the renormalization constant
connecting the lattice and continuum effective theories.
\end{abstract}


\newpage

\section{Introduction}
\label{sec:into}

The decay constant of the $B-$meson, $\fb$, defined through the
matrix element of the axial current, $A_\mu$,
\be
< 0 | A_\mu | B(p) > \equiv i f_B p_\mu
\label{eq:fb}
\ee
(where $B(p)$ is a $B-$meson of momentum $p$),
is an essential ingredient in many calculations in
the Standard Model of Particle Physics.
For example, it enters in the theoretical determinations of
(i) the mass splitting in the $B-\bar B$ system,
which is proportional to the square of $f_B$ and, in principle,
(ii) the Cabibbo-Kobayashi-Maskawa matrix element, $V_{ub}$, and hence,
through the unitary triangle to the element $V_{td}$ (see eg. \cite{rosner}).
Clearly therefore, for a more complete knowledge of some of
the fundamental parameters of the Standard Model, theorists need to
provide a prediction of $\fb$ (see also \cite{reina}).

The two main methods of determining $\fb$ from theory are QCD Sum Rules
(see eg. \cite{neubert}) and Lattice Gauge Theory (see eg.
\cite{sommer,bernard_dallas}). 
The QCD Sum Rules has an important advantage compared
to the lattice in that it is an analytic, rather than a numerical method.
However it involves approximations to the full theory that cannot
be systematically improved, and it has a perturbative component
calculated at a momentum scale $\mu \approx 1$ GeV
where the validity of perturbation theory can be questioned.

In contrast to QCD Sum Rules, the lattice is a fully non-perturbative
approach and furthermore uses approximations
that can be systematically improved.
For each approximation there is an associated {\em tunable}
``lattice parameter''
which can eventually be adjusted towards its physical value
thereby removing the approximation.
For example, lattice calculations are performed on
a finite physical volume which can, in principle,
be increased towards the physical value (ie. infinity)
such that finite volume effects can be neglected to
within any desired accuracy.
In general the size of the systematic errors introduced
due to these lattice approximations has been quantified
and often does not present a problem given
the size of the statistical errors.
This is particularly true in the following analysis.

A lattice approximation that warrants further discussion
is the ``Quenched Approximation''
where the effects of sea quark loops are suppressed.
All the lattice results discussed here were obtained
using the quenched approximation.
There is, however, strong circumstantial
evidence that its effects on hadronic masses are small
and its effects on hadronic matrix elements are of the level
of the statistical errors (10\% -- 20\%) \cite{weingarten_new}.
There will be a further discussion on quenching in the final section.

The reason that these lattice parameters are not set to their
physical values in the first place is due to the limitations
of present computing power.
In this sense, with the inevitable development of computing
resources, lattice QCD results will be able to provide predictions
with ever increasing precision.

Turning to the determination of $\fb$,
there is one immediate, technical problem in simulating a
$b-$quark on the lattice:
its mass is larger than the ultraviolet cut-off provided
by the finite spatial resolution of the lattice, ie. $m_b > \ia$.
Here $\ia$, the inverse lattice spacing, is typically of order
1--3 GeV in present simulations.
There are two approaches to overcome this problem.
One is to study quarks which are fairly heavy, but whose
masses are still less than $\ia$ and then to extrapolate
to $m_b$, and the second is to use
the ``static approximation'' which is the leading term
in the heavy quark expansion \cite{hqet}
\footnote{The static approximation will be discussed
in more detail in the next section.}.
This paper deals with lattice results using this second approach.
To obtain the real value of $\fb$
(ie. to all orders in the heavy quark expansion), one must
interpolate results from the two approaches \cite{abada}.

Historically, lattice calculations of $\fbs$ (ie. $\fb$ in the
static approximation), have proved very difficult due
to the low signal to noise ratio of the hadronic correlators
required for the calculation \cite{boucaud}.
The first successful measurements \cite{elc_60,wup1}
produced surprisingly large values, $\fbs \approx 300$ MeV.
Since then, many calculations, {\em at the same value of $a$},
have generally confirmed these early measurements
\cite{fermilab}--\cite{ape_sme0}.
However when calculations at different values of $a$ were
performed there appeared, in some analyses, to be a trend
towards smaller values
of $\fbs$ as $a\rightarrow 0$, ie. as the ``lattice parameter'', $a$,
is tuned towards its physical value
\cite{bernard_dallas,fermilab,wup2}.
The requirement for lattice results at a non-zero value
of $a$ to be physically relevant is that physical quantities
remain a constant in the continuum limit, $a \rightarrow 0$,
ie. that they {\em scale}
\footnote{This nomenclature is in analogy with
critical phenomena where all physical quantities are proportional
to the correlation length raised to a scaling index.}.
This apparent downward trend of $\fbs$, if actually present,
is therefore a violation
of scaling and is to be contrasted with the apparent 
scaling behaviour observed in other physical quantities \cite{doug}.
It must be stressed that, statements regarding scaling or the lack of it
can only be made within a certain statistical uncertainty,
and for quantities as difficult to determine as $\fbs$,
these errors are typically large. At some point scaling violations
may disappear below the level of statistics, and hence become unimportant.

The issue of scaling is normally studied by plotting the
ratios of physical quantities as a function of $a$,
as was the case in the above-mentioned analyses of $\fbs$
in which an apparent violation of scaling was observed.
This paper outlines an alternative method of studying scaling
behaviour in which the functional form of the dimensionless lattice
quantity corresponding to $\fbs$, (ie. $\zl$, defined later in
eq.(\ref{eq:zl_define})),
is fitted as a function of the bare lattice coupling.
We show that this analysis applied to the currently published data
on $\fbs$ shows that its scaling violations for $\ia \gtap 2$ GeV
are small, ie. that they have fallen to below the statistical errors.
However, for $\ia \ltap 2$ GeV we do find scaling violations.
This suggests that the conclusions of the studies
\cite{bernard_dallas,fermilab,wup2} are due to the inclusion of
data too far from the continuum, ie. with $\ia < 2$ GeV.
The implication of this result is that, at present, the best
estimate for the continuum value of $\fbs$ is given by those
simulations at $\ia \approx 2-3$ GeV.

In the following section we outline the static theory as applied
on the lattice. Those readers familar with the static theory
on the lattice may wish to skip this section. In sec.3 we study the scaling
of $\fbs$ and other physical quantities,
and in sec.4 we discuss our results.
We also investigate finite lattice spacing, or $O(a)$, effects
and the approach to asymptotic scaling.
Once these issues are studied and understood we obtain a continuum value
for $\fbs$ of $230(35)$ MeV.

\section{Static theory on the lattice}

This section begins with a brief description of the
lattice method of calculating physical quantities.
For a full discussion see, for example, the review articles
\cite{sommer,gm}.

Lattice QCD typically studies Euclidean correlation functions, $C(t)$,
of hadronic operators $O^L$,
\be
C(t) = < O^L(t) \; O^L(0) >.
\label{eq:ct1}
\ee
For purposes of explanation, we will define
\be
O^L(t) = \sum_{\tilde{x}} A_0(\tilde{x},t)
       = \sum_{\tilde{x}} \psi_1(\tilde{x},t) \gamma_0 \gamma_5
                          \psi_2(\tilde{x},t),
\ee
where the $\gamma_{0,5}$ are Dirac spinor matrices, and the subscripts
$1,2$ are flavour indices.
At large times $t$, $C(t)$ has the following behaviour:
\be
C(t) \rightarrow \frac{|< 0 | O^L | P >|^2}{2M^L_P} e^{-M^L_P \; t}
\label{eq:ct}
\ee
where $P$ is the lowest state with the quantum numbers
of $O^L$, in our case a pseudo-scalar meson.
Once the exponential behaviour in eq.(\ref{eq:ct}) has been
established, a simple exponential fit provides lattice predictions
of the decay constant,
\be
f_P^L = <0|O^L|P> / M_P^L,
\ee
(cf. eq.(\ref{eq:fb})) and the mass, $M^L_P$.
We will refer to these lattice predictions, $f_P^L$ and $M_P^L$,
generically as $\Omega^L_i$
(whether they be hadron masses or matrix elements).
These are not physical quantities because
(i) all the fields in the lattice action are made dimensionless
using the lattice spacing $a$, and therefore all
the predictions, $\Omega^L_i$, from the lattice are also dimensionless, and,
(ii) in the case of matrix elements, the operator $O^L$,
is not correctly normalized.

To determine the scale $a$ we choose a
lattice prediction of {\em one} physical quantity and compare it to
the experimental (dimensionful) value, $\Omega_i$.
The lattice operator $O^L$ is correctly normalized to its continuum
counterpart $O$ typically through a multiplicative renormalization factor
$Z^{Ren}$, ie. $O = Z^{Ren} \; O^L$ (see eg. \cite{sach}).
Combining both steps we have:
\be
\ia_i = \frac{\Omega_i}{Z_i^{Ren} \; \Omega_i^L}.
\label{eq:set}
\ee
(In this discussion we are assuming that $\Omega_i$ has energy
dimension unity.) For non-matrix element quantities such as
hadron masses $M^L$, $Z^{Ren} \equiv 1$.

In lattice QCD, we have dimensional transmutation in action:
we first set the bare coupling in the lattice action,
$g^2$, at the start of the calculation, and then determine the
corresponding ultraviolet cutoff, $\ia$, through eq.(\ref{eq:set}).
(We could instead proceed in the reverse direction, by first setting
$\ia$, but this would require expensive simulations at many trial values
of $g^2$ before we settled at the correct value of $g^2$ corresponding
to the chosen value of $\ia$.)

Finally then we have the lattice prediction of a dimensionful
physical quantity, $\Omega_j$:
\bea
\Omega_j   & = & Z_j^{Ren} \; \Omega_j^L \; \ia_i \label{eq:omega} \\
           & = & \Omega_i \; \frac{Z_j^{Ren} \; \Omega_j^L}
                                {Z_i^{Ren} \; \Omega_i^L}
\nonumber \eea

The correlation function $C(t)$ is numerically calculated using the
Wick contracted form of eq.(\ref{eq:ct1}),
\be
C(t)=< Tr\{ G_1(t,0) \gamma_0 \gamma_5 G_2(0,t) \gamma_0 \gamma_5 \}>.
\ee
The quark propagators $G_{1,2}$ are defined from the lattice
version of the Dirac equation.

In the static version of the theory $G_1$ is defined using the
solution of the Dirac equation in the limit of an infinitely
heavy quark (see eq.(2.37) in \cite{sommer}).
In this case we are simulating a pseudo-scalar meson (which we'll
denote $B$) made up of a light quark and an infinitely heavy quark.
We have from eqs.(\ref{eq:fb},\ref{eq:ct}),
\be
C(t) \rightarrow \frac{(f_B^{stat,L})^2 \; M_B \; a}{2} e^{-E_B \; t},
\ee
where $E_B$ is the binding energy\footnote{
We have factored out the exponential time dependence
of the (infinitely-heavy) quark in the static propagator.}
of the light quark in the B-meson,
and $M_B$ is the experimental value of the $B-$meson mass.
We define
\be
Z_L^2 \equiv (\Omega^L_{f_B})^2 = \frac{(f_B^{stat,L})^2 \; M_B \; a}{2}.
\label{eq:zl_define}
\ee
This is the lattice quantity corresponding to $\fbs$ (see sec.1).
Note that in \cite{fermilab} the symbol $\tilde{f_B}$ is used for $\zl$.
Correctly normalizing the operator, and putting in the appropriate dimensions
(cf. eq.(\ref{eq:omega})) we have,
\be
\fbs = \sqrt{\frac{2}{M_B}} \; Z^{Ren}_{f_B} \; Z_L \; a^{-3/2}.
\label{eq:fbs_define}
\ee
The issue of scaling is then simply a question of whether
the $g^2$ dependences of the $Z^{Ren}_{f_B}, Z_L$ and $a^{-3/2}$ cancel
in eq.(\ref{eq:fbs_define}).

Using eq.(\ref{eq:set}) we have finally
\be
\fbs = \sqrt{\frac{2 \; \Omega_i^3}{M_B}} \;
\frac{Z^{Ren}_{f_B} \; Z_L}{(Z_i^{Ren} \; \Omega_i^L)^{3/2}}.
\label{eq:fbs_rat}
\ee

\section{Continuum scaling of $\fbs$}

The usual method of determining whether the scaling\footnote{
as opposed to {\em asymptotic} scaling - see sec.4} region of
a lattice simulation has been reached is to study a dimensionless
ratio, $\Omega_1 \over \Omega_2$, of physical quantities $\Omega_i$,
to see if this ratio is a constant in $a$, or equivalently in $g^2$.
Using eq.(\ref{eq:omega}) the $\Omega_i$ in the ratio can be replaced
with the dimensionless quantity $Z^{Ren}_i \Omega_i^L$.
Dimensionless ratios of lattice quantities are generally used
wherever possible because statistical fluctuations, and indeed,
some systematic effects tend to cancel. For this reason ratios are often
better determined than the absolute quantities themselves.

\begin{figure}[t]   
  \begin{center}
  \leavevmode
  \epsfysize=540pt
  \epsfbox{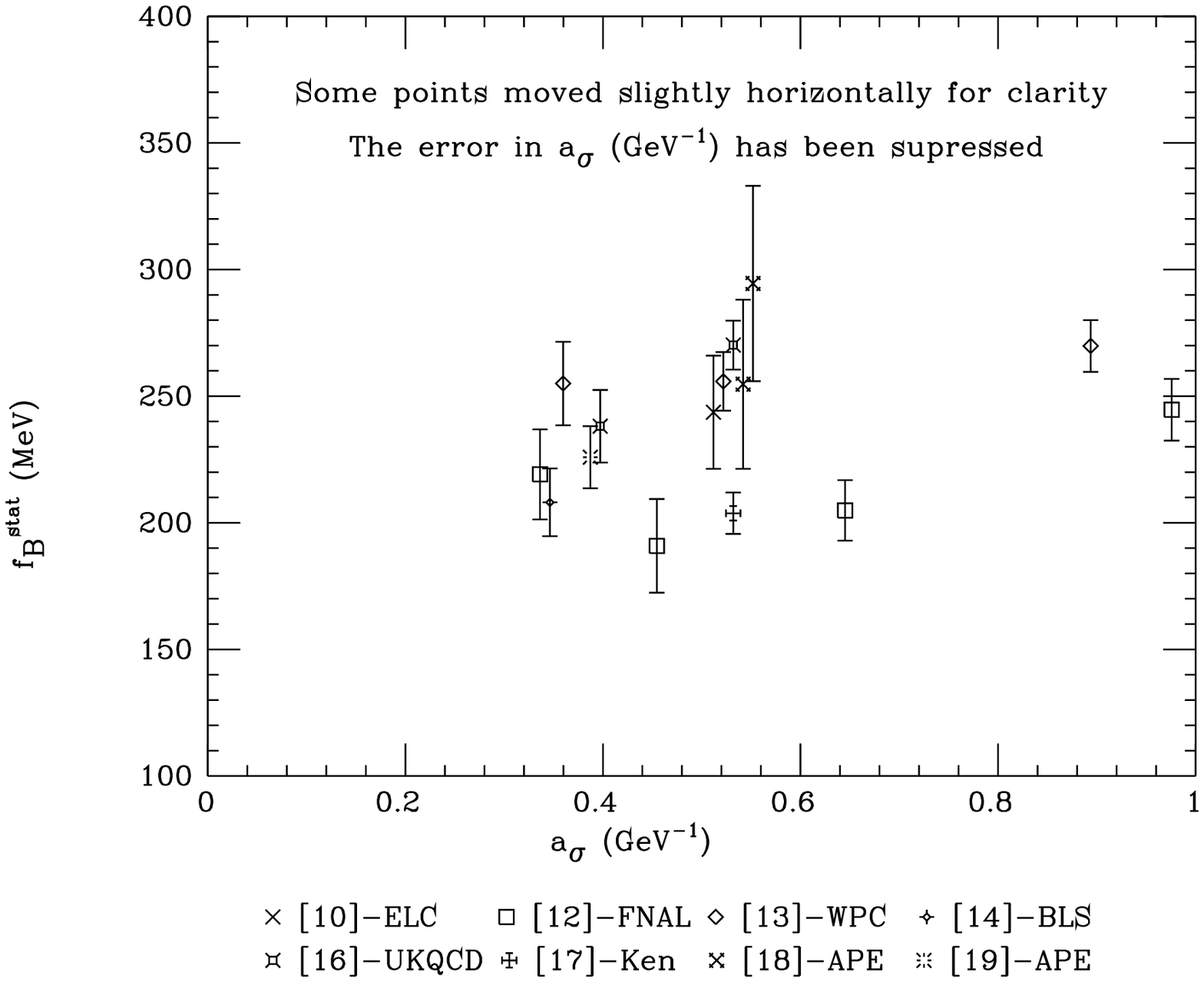}
  \end{center}
  
  \caption[]{\it{Plot of $\fbs$ against the lattice spacing as determined
  from the string tension, $\ia_\sigma$, from various groups as listed in
  the legend.
  $\fbs$ has been determined using eq.(\ref{eq:fbs_define}) with $\ia = 
  \ia_\sigma$.
  See text for the definition of $Z^{Ren}_{f_B}$ used.}}
      \protect\label{fig:bernard}
  \end{figure}

Using this ``ratio'' method, we show in fig.(\ref{fig:bernard})
the plot of $\fbs$ versus $a_\sigma$, ie. the lattice
spacing determined from the string tension $\sigma$.
The data plotted is a collection of published data on $\fbs$
and is reported in table \ref{tab:zl} with references cited in column 1.
(Note that $\beta = 6/g^2$, where $g$ is the bare lattice coupling
which appears in the lattice action.)
For completeness, in table \ref{tab:zl}, we also list the values
of $Z^{Ren}_{f_B}$ and $\ia$ used and the $\fbs$ obtained by each group.
In the figure we use eq.(\ref{eq:fbs_rat}) to determine the $\fbs$ values
given the $Z_L$ values in table \ref{tab:zl}.
The $Z^{Ren}_{f_B}$ values we have used are the
boosted, tadpole improved values \cite{lepage,tadpoles},
which means, for instance, at $\beta = 6.0$ we have $Z^{Ren}_{f_B} = 0.70$
for the Wilson action, and $0.79$ for the clover action.

It is difficult to interpret this plot, and difficult to gauge
the likelyhood of a decrease or increase of $\fbs$ as $a \rightarrow 0$.
Of course, one can simply use a linear or quadratic fit
\cite{sommer,bernard_dallas,fermilab,wup2}, and fit the data in any case.
The problem with this approach is that continuum value of $\fbs$
in fig.(\ref{fig:bernard}) is the extrapolation of the ratio of two quantities
(in this case $Z_L$ and $(\Omega^L_\sigma)^{3/2}$, see eq.(\ref{eq:fbs_rat}))
which have {\em very similar} functional dependencies on $a$.
Any slight systematic effect in either the numerator or denominator
will swing the ratio, which therefore could significantly affect the
extrapolation.

In this paper we show that such a systematic effect is present
in the form of scaling violations for $\ia \ltap 2$ GeV.
In the approach presented here, the functional dependence of
the numerator and denominator are determined separately,
and then compared to check for scaling.
We then present a method which does not suffer from the above problem
and enables a continuum estimate of $\fbs$ to be made.

  \begin{table}
  \begin{center}
  \begin{tabular}{lllllll}
  \hline        
  \hline
  Ref. & 
  Action & $\beta=6/g^2$ & $Z_L$ & $a^{-1}$  & $Z^{Ren}$ &$f_B^{stat}\;$MeV\\
  \hline
  \\
  \cite{elc_60} - ELC &
  Wilson & 6.0     & 0.22(2)   & $2.0(2)^b$  & $0.8^g$   & 310(25)(50) \\
  \\
  \cite{wup2} - WPC &
  Wilson & 5.74    & 0.543(20) & $1.119(8)^e$&           &             \\
  \cite{wup2} - WPC &
  Wilson & 6.0     & 0.231(10) & $1.88(2)^e$ &           &             \\
  \cite{wup2} - WPC &
  Wilson & 6.26    & 0.125(8)  & $2.78(2)^e$ &           &             \\
  \\
  \cite{fermilab} - FNAL &
  Wilson & 5.7     & 0.564(28) & $1.15(8)^f$ & $0.63^h$  & 271(13)(20) \\
  \cite{fermilab} - FNAL &
  Wilson & 5.9     & 0.250(14) & $1.78(9)^f$ & $0.65^h$  & 241(13)(13) \\
  \cite{fermilab} - FNAL &
  Wilson & 6.1     & 0.135(13) & $2.43(15)^f$& $0.68^h$  & 215(21)(14) \\
  \cite{fermilab} - FNAL &
  Wilson & 6.3     & 0.099(8)  & $3.08(18)^c$& $0.68^h$  & 225(17)(14) \\
  \\
  \cite{BLS} - BLS &
  Wilson & 6.3     & 0.094(6)  &$3.21(9)(17)^d$& $0.7^h$ & 235(20)(21) \\
  \\
  \cite{ukqcd} - UKQCD &
  clover & 6.0     &$0.211^{+6}_{-7}$&$2.0^{+3b}_{-2}$   &
  $0.78^i$ & $286^{+8+67}_{-10-42}$ \\
  \cite{ukqcd} -UKQCD &
  clover & 6.2     &$0.117^{+7}_{-7}$&$2.7^{+7b}_{-1}$   &
  $0.79^i$ & $253^{+16+105}_{-15-14}$ \\
  \\
  \cite{ken} - Ken &
  Wilson & 6.0     & $0.184(7)^a$ & 2.0    & $0.70^h$ & $224^{+9}_{-7}$ \\
  \\
  \cite{ape_6018} - APE &
  Wilson & 6.0     & 0.23(3)   & $2.11(5)(10)^b$ & $0.8^g$ & 350(40)(30) \\
  \cite{ape_6018} - APE &
  clover & 6.0     & 0.23(3)   & $2.05(6)^b$     & $0.89^g$& 370(40)     \\
  \\
  \cite{ape_sme0} - APE &
  clover & 6.2     & 0.111(6)  & $3.0(3)^b$      & $0.81^h$& 290(15)(45) \\
  \hline
  \hline
  
  \end{tabular}
  \caption{ {\it Values for $Z_L$ and $\fbs$ from various group's work.}
  \newline
  $^a$ this $Z_L$ value was obtained from eq.(\ref{eq:fbs_define}),
  ie. the $Z_L$ value was not explicitly published
  \newline
  $^b$ $a^{-1}$ from averaging scale obtained from $f_\pi$ and $M_\rho$
  \newline
  $^c$ $a^{-1}$ from the 1P-1S value at $\beta = 6.1$ using 1-loop
  asymptotic freedom to extrapolate to $\beta=6.3$
  \newline
  $^d$ $a^{-1}$ from $f_\pi$
  \newline
  $^e$ $a^{-1}$ from the string tension, $\sigma$
  \newline
  $^f$ $a^{-1}$ from $1P-1S$ charmonium splitting
  \newline
  $^g$ $Z^{Ren}$ from standard perturbative result
  \newline
  $^h$ $Z^{Ren}$ from boosted, tadpole improved analysis
  \newline
  $^i$ $Z^{Ren}$ from boosted analysis}
  \label{tab:zl}
  \end{center}
  \end{table}

\begin{figure}[t]   
  \begin{center}
  \leavevmode
  \epsfysize=540pt
  \epsfbox{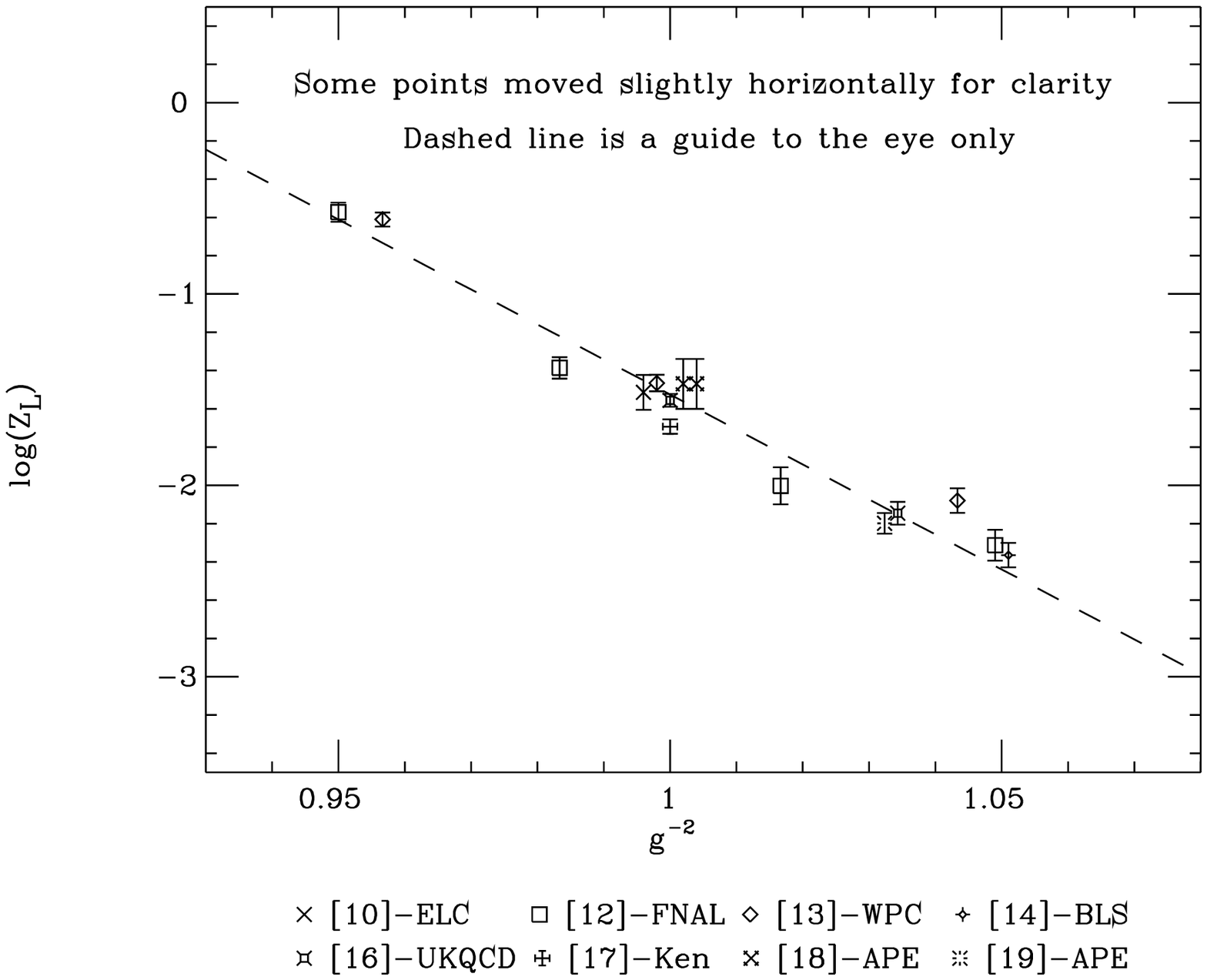}
  \end{center}
  
  \caption[]{\it{Plot of $log(Z_L)$ from various groups as listed in the legend.
  $g$ is the bare lattice coupling.}}
      \protect\label{fig:zl}
  \end{figure}

We begin by assuming a naive scaling of the $\Omega^L$ quantity
appropriate for $\fbs$, ie. $\zl$ (see eq.(\ref{eq:zl_define}))
\be
\zl \sim e^{-9S_{f_B}/g^2}.
\label{eq:zl}
\ee
We will justify this choice of functional behaviour in the next
section and show that, for our purposes, there is no loss of
generality in eq.(\ref{eq:zl}).
Assuming this relationship, we now plot in
fig.(\ref{fig:zl}) $log(\zl)$ against $1/g^2$.
Later we will fit this plot to a straight line to extract $S_{f_B}$
\footnote{Note that for ease of presentation, we have plotted
$\zl$ values obtained with both Wilson and clover \cite{sw}
actions together on the same plot. A priori, we cannot expect the same
value of $S_{f_B}$ from both actions, so in the determinations
of $S_{f_B}$ we fit results from the two actions separately.}.
We notice immediately from fig.(\ref{fig:zl}) that the
data fall in a roughly linear band,
and with a relatively small spread.
One could imagine using this plot to check future calculations
of $\zl$ at smaller $a$ values.
It is obvious that the behaviour of $\zl$ with $g^2$
apparent in fig.(\ref{fig:zl}) is more clearly manifest
than the behaviour of $\fbs$ with $a$ (see fig.(\ref{fig:bernard})).

We now determine $S_{f_B}$ from the slope of $\;log(Z_L)\;$
against $1/g^2$.
For the Wilson data, we choose two
intervals: $5.7 \le \beta \le 6.0$ and $6.0 \le \beta \le 6.3$.
The motivation for these two intervals is that there is evidently
a dependence of $S_{f_B}$ on $g^2$. This will be discussed
in detail in the next section. We obtain the values
reported in the third column of table \ref{tab:s}.
For the clover data, we choose the interval $6.0 \le \beta \le 6.2$
since the $\fbs$ data is available only at $\beta = 6.0$ \& $6.2$.

  \begin{table}
  \begin{center}
  \begin{tabular}{lcccccc}
  \hline
  \hline
  \it{Action} & \it{Fitting interval} &
  $S_{f_B}$ & $S_\sigma$ & $S_{M_\rho}$ & $S_{f_\pi}$ & $S_{1P-1S}$ \\
  \hline
  \\
  Wilson  & $5.7 \le \beta \le 6.0$ &
  2.39(10)& 2.06(3) & 1.53(7)  & 1.8(2) & 2.2(4) \\
          & $6.0 \le \beta \le 6.3$ &
  1.57(11)& 1.52(6) & 1.35(13) & 1.7(2) & ---    \\
  \\
  \hline
  \\
  Clover & $6.0 \le \beta \le 6.2$ &
  2.1(2)  & 1.52(6) & 1.7(2)  & 2.1(3) & --- \\
  \\
  \hline
  \hline
  \end{tabular}
  \caption{ \it{Values for $S_i$ obtained from fits to the data in
  tables \ref{tab:zl}, \ref{tab:a_wil} and \ref{tab:a_clo}
  to eqs.(\ref{eq:zl},\ref{eq:a}). Note $\beta = 6/g^2$.}}
  \label{tab:s}
  \end{center}
  \end{table}
  \begin{table}
  \begin{center}
  \begin{tabular}{lllllll}
  \hline
  \hline
  Ref. & 
  $\beta=6/g^2$  & $a^{-1}_\sigma$ & $a^{-1}_{M_\rho}$&$a^{-1}_{f_\pi}$&
  $Z^{Ren}_{f_\pi}$ & $a^{-1}_{1P-1S}$ \\
  \hline
  \cite{wup2} - WPC &
  5.7      & 1.025(3)    &             &           &      & \\
  \cite{aida} - FNAL &
  5.7      &             &             &           &      & 1.15(8) \\
  \cite{weingarten_new} - GF11 &
  5.7      &             & 1.42(2)     & 1.25(5)   & 0.75 & \\
  \cite{wup2} - WPC &
  5.74     &             & 1.44(3)     &           &      & \\
  \cite{wup2} - WPC &
  5.8      & 1.272(6)    &             &           &      & \\
  \cite{wup2} - WPC &
  5.9      & 1.55(2)     &             &           &      & \\
  \cite{aida} - FNAL &
  5.9      &             &             &           &      & 1.78(9) \\
  \cite{weingarten_new} - GF11 &
  5.93     &             & 1.99(4)     & 2.00(5)   & 0.77 & \\
  \cite{wup2} - WPC &
  6.0      & 1.88(2)     & 2.25(10)    &           &      & \\
  \cite{ape_6018} - APE &
  6.0      &             & 2.23(5)     & 2.21(8)   & 0.78 & \\
  \cite{ape_dallas} - APE &
  6.0      &             & 2.18(9)     & 1.97(8)   & 0.78 & \\
  \cite{aida} - FNAL &
  6.1      &             &             &           &      & 2.43(15) \\
  \cite{weingarten_new} - GF11 &
  6.17     &             & 2.77(4)     & 2.82(7)   & 0.79 & \\
  \cite{wup2} - WPC &
  6.2      & 2.55(1)     &             &           &      & \\
  \cite{ape_dallas} - APE &
  6.2      &             & 2.88(24)    & 2.96(24)  & 0.79 & \\
  \cite{wup2} - WPC &
  6.26     &             & 3.69(32)    &           &      & \\
  \cite{wup2} - WPC &
  6.4      & 3.38(1)     &             &           &      & \\
  \cite{abada} - ELC&
  6.4      &             & 3.70(15)    & 4.0(6)    & 0.80 & \\
  \hline
  \hline
  \end{tabular}
  \caption{ \it{Values for $a^{-1}$ obtained from various group's work
  using the Wilson action.}}
  \label{tab:a_wil}
  
  \begin{tabular}{lllll}
  \hline
  \hline
  Ref. & 
  $\beta=6/g^2$  & $a^{-1}_{M_\rho}$  & $a^{-1}_{f_\pi}$ &
  $Z^{Ren}_{f_\pi}$ \\
  \hline
  \cite{ape_6018} - APE &
  6.0      & 2.05(6)         & 2.11(11)      & 1.09(3) \\
  \cite{ape_semi} - APE &
  6.0      & 1.92(11)        & 1.94(5)       & 1.09(3) \\
  \cite{ape_no60} - APE &
  6.0      & 1.95(7)         & 1.78(9)       & 1.09(3) \\
  \cite{ukqcd} - UKQCD &
  6.2      & 2.7(1)          & 3.2(2)        & 1.04(1) \\
  \cite{ape_sme0} - APE &
  6.2      & 3.05(19)        & 2.73(17)      & 1.04(1) \\
  \hline
  \hline
  \end{tabular}
  \caption{ \it{Values for $a^{-1}$ obtained from various group's work
  using the Clover action.}}
  \label{tab:a_clo}
  \end{center}
  \end{table}


Of course, the values of $\zl$ and their scaling with $g^2$ 
are not enough to determine the scaling of $\fbs$ itself.
One needs to also study the $g^2$ behaviour of some $\Omega_i^L$
in order to set the scale.
Four such choices are $\sqrt\sigma, \mr, \fpi$, and the $1P-1S$
splitting in charmonium.
In tables \ref{tab:a_wil} \& \ref{tab:a_clo}, a sample of the
published data on $\ia$ from various collaborations
is presented with the references cited in column 1.
Note that for $a^{-1}_{f_\pi}$, we have used
the renormalization constants shown in the tables.
Specifically these were obtained with the
``boosted'' perturbation theory\footnote{
In this case we choose the ``boosted'' coupling $g_V^2 = g^2 / u_0^4$
where $u_0^4$ is the average plaquette, see \cite{lepage}.}
in the Wilson case
\cite{lepage}, and the non-perturbatively obtained values
in the clover case \cite{vlad1,vlad2}.

Again we assume the data to have the
following functional form (see eg. \cite{abada}):
\be
a_i = \frac{Z_i^{Ren} \; \Omega_i^L}{\Omega_i} \sim e^{-6 S_i / g^2}
\label{eq:a}
\ee
where $i = {\sigma,M_\rho,f_\pi,1P-1S}$, and we have used eq.(\ref{eq:set}).
As in eq.(\ref{eq:zl}), this functional form can be assumed
with no loss of generality.

\begin{figure}[t]   
  \begin{center}
  \leavevmode
  \epsfysize=540pt
  \epsfbox{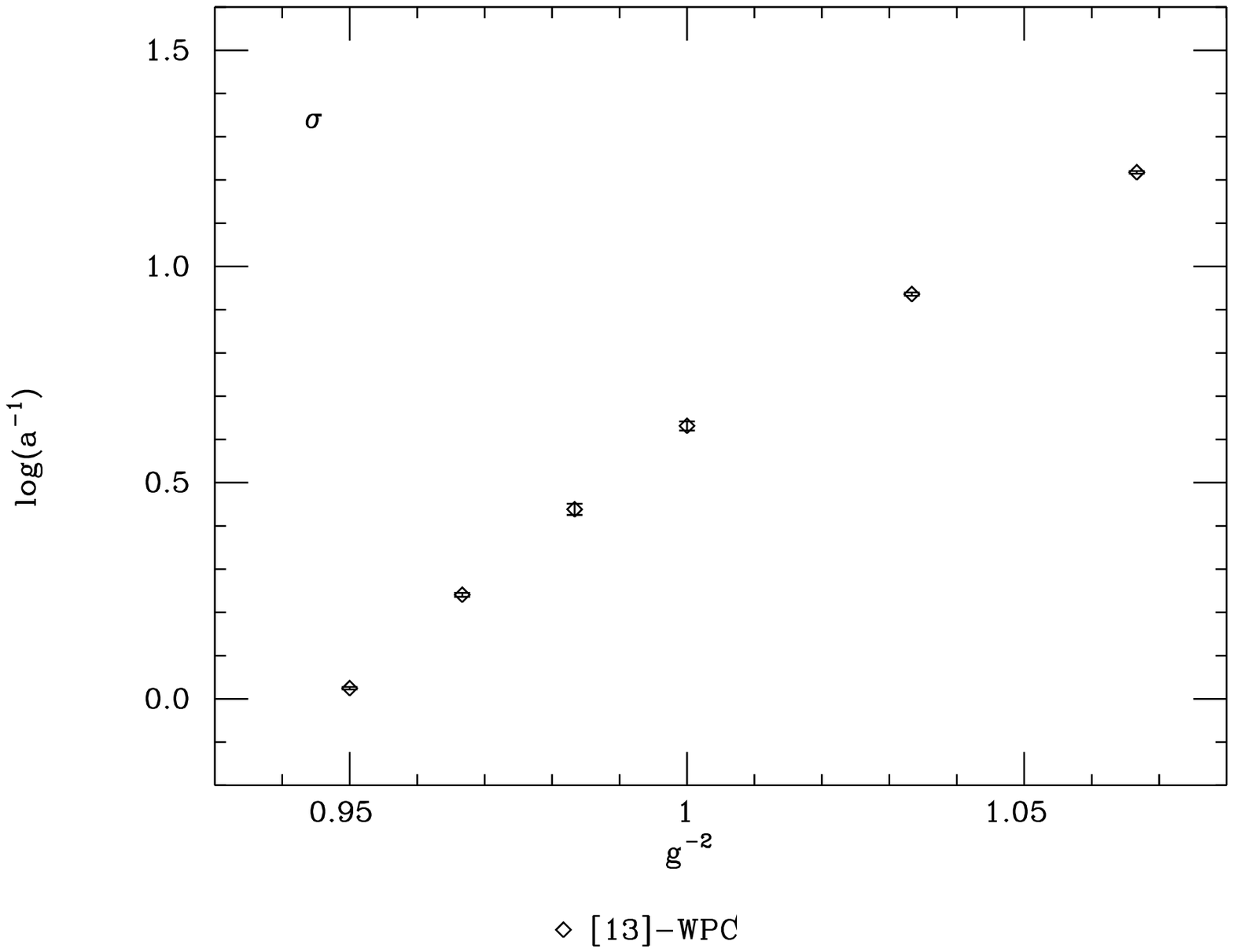}
  \end{center}
  
  \caption[]{\it{Plot of $log(\ia)$ from the string tension.
  The reference is as appears in the legend.}}
      \protect\label{fig:aa}
  \end{figure}

\begin{figure}[t]   
  \begin{center}
  \leavevmode
  \epsfysize=540pt
  \epsfbox{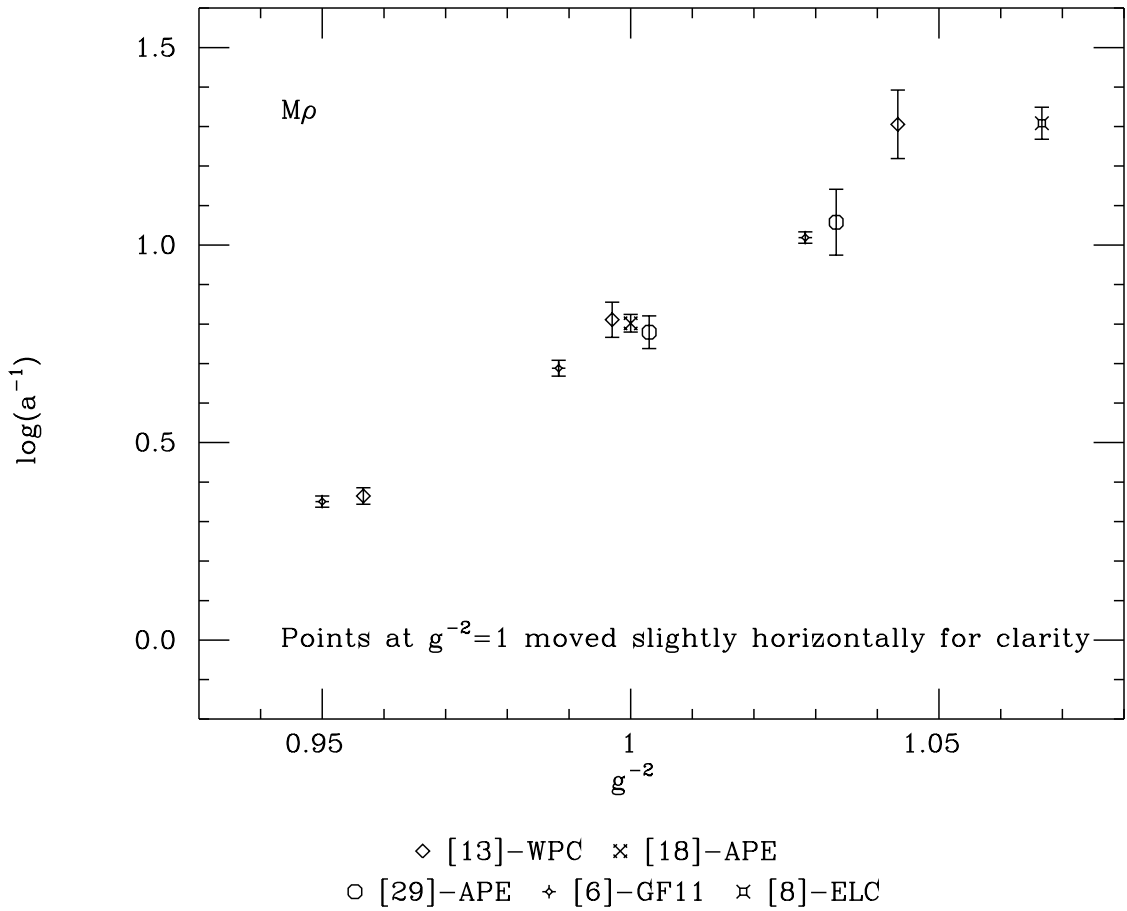}
  \end{center}
  
  \caption[]{\it{Plot of $log(\ia)$ from $M_\rho$ for the Wilson action.
  The references are as appears in the legend.}}
      \protect\label{fig:ab}
  \end{figure}

\begin{figure}[t]   
  \begin{center}
  \leavevmode
  \epsfysize=540pt
  \epsfbox{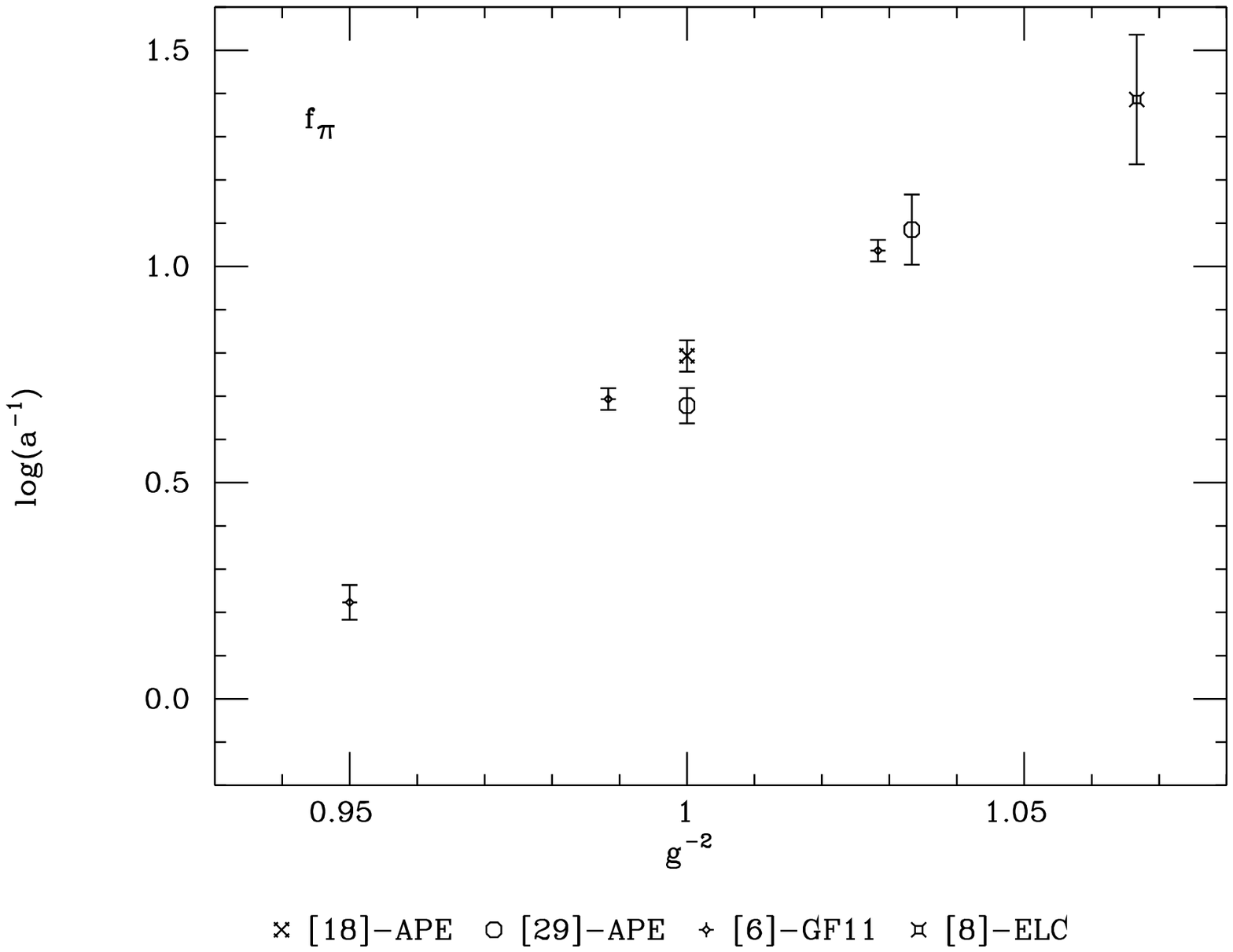}
  \end{center}
  
  \caption[]{\it{Plot of $log(\ia)$ from $f_\pi$ for the Wilson action.
  The references are as appears in the legend.}}
      \protect\label{fig:ac}
  \end{figure}

\begin{figure}[t]   
  \begin{center}
  \leavevmode
  \epsfysize=540pt
  \epsfbox{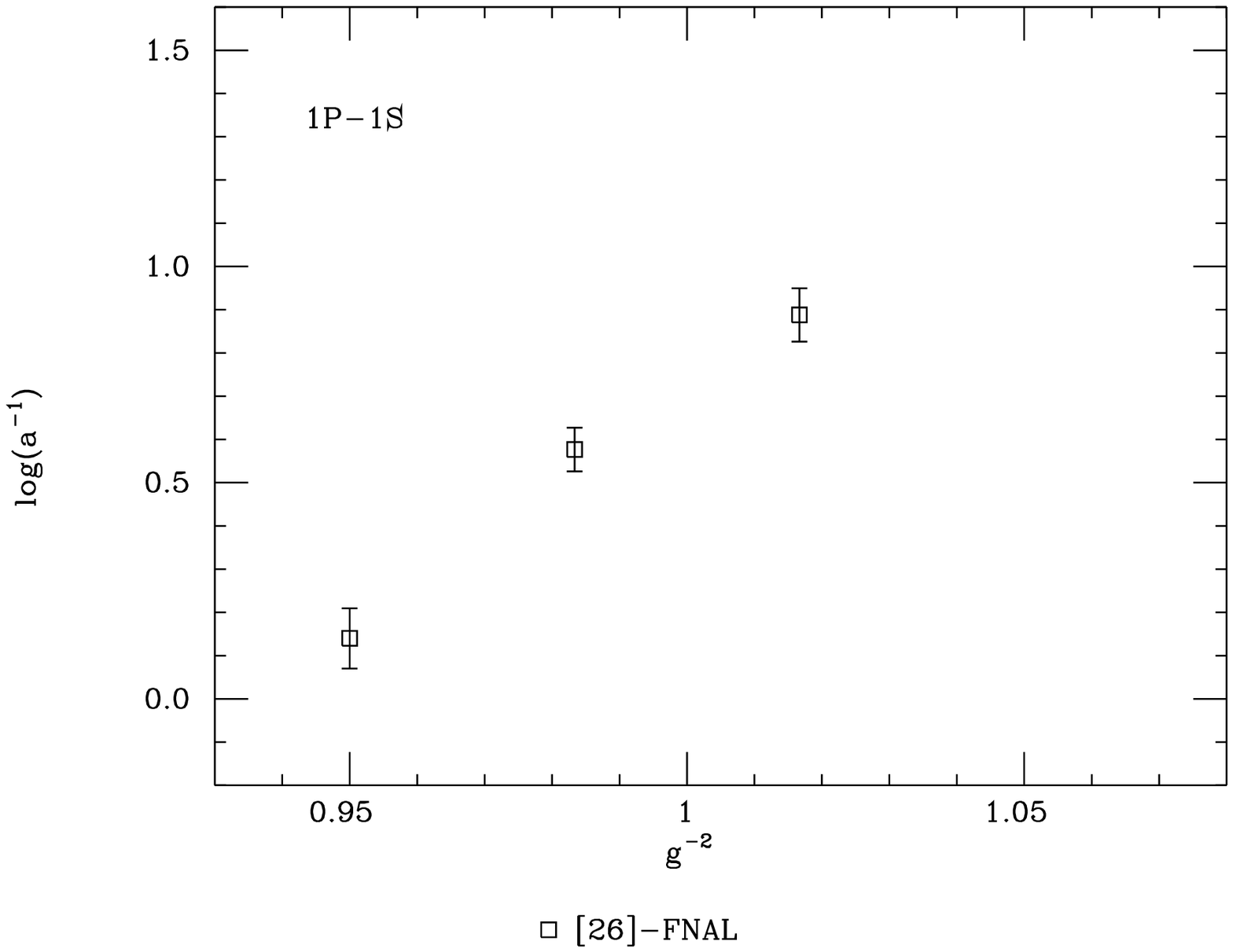}
  \end{center}
  
  \caption[]{\it{Plot of $log(\ia)$ from the $1P-1S$ splitting in charmonium
  for the Wilson action. The reference is as appears in the legend.}}
      \protect\label{fig:ad}
  \end{figure}

In figs.(\ref{fig:aa}-\ref{fig:ad}), $log(a_i)$ is plotted against $1/g^2$
for the Wilson data from table \ref{tab:a_wil}.
From the gradient of this plot fitted in the same intervals
as the fit of $log(\zl)$, we obtain the values of $S_i$,
$i=\{\sigma,M_\rho,f_\pi,1P-1S\}$ reported in table \ref{tab:s}.

We now turn to a discussion of the continuum limit of $\fbs$.
The functional behaviour of $\fbs$,
using eqs.(\ref{eq:fbs_define},\ref{eq:zl},\ref{eq:a}) is:
\be
\fbs(g^2) \sim  \frac{Z_L}{a_i^{3/2}} = e^{-9(S_{f_B} - S_i)/g^2}.
\label{eq:ffbb}
\ee
We have ignored the $g^2$ dependence of the renormalization
constant $Z^{Ren}_{f_B}$, and will justify this below.
Thus, the issue of scaling is addressed in this analysis
by a comparison of $S_{f_B}$ and $S_i$.
We believe this is a cleaner method of studying
the scaling of $\fbs$ since the systematics present
in both $Z_L$ and $\ia_i$ can be isolated and studied.

From table \ref{tab:s} for $\beta=6/g^2 \le 6.0$ there is a clear statistical
evidence for a {\em violation} of scaling in the Wilson data,
ie. the $S_i$ are not all compatible.

For $6.0 \le \beta \le 6.3$, in the Wilson case, all the $S_i$ are
within around one standard deviation of the value $1.5$.
In fact both the $S_\sigma$ and $S_{f_\pi}$ values in table \ref{tab:s}
are probably a little too high because there is no
$\ia_\sigma$ or $\ia_{f_\pi}$ data in this range above $\beta=6.2$.
The conclusion is therefore that
$\fbs, \sigma, M_\rho, f_\pi$ and $1P-1S$ mutually scale within present
statistical errors for $\ia \gtap 2$ GeV.
This is the main result of this paper.
Results which suggest a decrease of $\fbs$ as $a \rightarrow 0$ can
now be understood to be due to the inclusion of $\beta < 6.0$ data
which suffer from systematic effects due to the lack of scaling
of the quantities involved.

In the clover case, there exists
only two collaboration's results, and at only two $\beta$ values,
and so an interpretation of the results, in this case,
may be premature.
To elaborate, if the UKQCD results \cite{ukqcd} are not included
in the determination of the $S_i$, we obtain the following values:
\bea
S_{f_B} = 2.1(2), & S_{M_\rho} = 2.1(3), & S_{f_\pi} = 1.7(3),
\eea
to be compared with the values in the last row of table \ref{tab:s}.
Evidently the $S_{M_\rho}$ and $S_{f_\pi}$ values interchange when the
UKQCD results are not included. This suggests that more
results, particularly at larger $\beta$ are required to settle
the $S_i$ values in the clover case.
In any case, the results suggest that
$\fbs, M_\rho$ and $f_\pi$ mutually scale also in the clover case
(for $\beta \gtap 6.0$).

\section{Discussion}


In the rest of the paper asymptotic scaling, lattice artefacts
and the continuum value of $\fbs$ will be discussed.

Asymptotic scaling is where the $g^2$
dependence of lattice quantities $\Omega^L_i$ is the same as that
predicted by weak-coupling perturbation theory.
This behaviour can be obtained by integrating the
beta-function of the theory,
\be
\beta(g^2) = a \frac{dg^2}{da}
           = 2 \beta_0 \frac{g^4}{16\pi^2}
           + 2 \beta_1 \frac{g^6}{(16\pi^2)^2}
           + O(g^8),
\ee
to obtain,
\be
a = \Lambda^{-1} \;\; f_{PT}(g^2),
\label{eq:a_PT}
\ee
where
\be
f_{PT} = (g^2)^{-\beta_1 \over {2 \beta_0^2}} \;\;
         e^{- \frac{16\pi^2}{2\beta_0} \frac{1}{g^2} },
\ee
$\beta_{0,1}$ are given by:
\be
\beta_0 =\frac {(11N-2 n_f)} {3},  \,\,\,\,\,\,\,\,\,\,\,
\beta_1 = \frac {34}{3} N^2 - \frac{10}{3} N n_f-\frac{(N^2-1)}{N}n_f,
\ee
$n_f$ is the number of flavours, $N$ the number of colours
and $\Lambda$ is some constant of integration.
The subscript ``$PT$" refers quantities obtained from
(second-order) weak-coupling perturbation theory.
For the quenched theory\footnote{see eg. \cite{unq} for a discussion
of asymptotic scaling in the unquenched theory}, $n_f = 0$, so we have
from eq.(\ref{eq:a})
\be
S_{PT} = \frac{1}{6} \;\; \frac{16\pi^2}{22} \approx 1.20,
\label{eq:s_PT}
\ee
where we have ignored higher order terms (ie. set $\beta_1=0$).
We expect, for small enough $g^2$, that $S_i = S_{PT}$.
From table \ref{tab:s} we see that this is not the case.
In the following, we discuss the possible causes for
this discrepancy.

\subsection{Effects of higher order terms}

The first and most obvious explanation for the inequality,
$S_i \ne S_{PT}$, is that for the values of
$g^2$ in table \ref{tab:s}, the effect of higher order terms
($\beta_1, \beta_2,$ ...) is significant. 
Note moreover that the downward
trend of the $S_i$ towards $S_{PT}$ as $g^2$ decreases is at least consistent
with the prediction of weak-coupling perturbation theory
and the declining importance of the higher order corrections in this limit.

However, to study this hypothesis quantitatively, we first note that
from eq.(\ref{eq:fbs_define}), the lattice prediction of $\fbs$ is given by:
\be
\fbs \sim Z_{f_B}^{Ren}(g^2) \;\; Z_L(g^2) \;\;
     (g^2)^{\frac{3 \beta_1}{4 \beta_0^2}} \;\; e^{9 S_{PT} / g^2},
\label{eq:omegaPT}
\ee
where we have assumed the two-loop formula (eq.(\ref{eq:a_PT})) for $a$,
and have included a $g^2$ dependence in $Z^{Ren}$.
$Z_{f_B}^{Ren}$ has the following expansion in $g^2$:
\be
Z_{f_B}^{Ren}(g^2) = 1 - \epsilon g^2 + O(g^4),
\label{eq:z_ren}
\ee
where $\epsilon \approx 0.2$.
In this formula we have taken only the matching between the lattice
and continuum effective theories and have ignored the anomalous dimension
\cite{z_cont} (see also \cite{fermilab}).
It can easily be demonstrated that
this extra factor does not affect the following discussion.

Since $\fbs$ is a fixed physical number, we have, using eqs.(\ref{eq:zl},
\ref{eq:omegaPT} \& \ref{eq:z_ren})
\be
Z_L(g^2)
\sim
( 1 + \epsilon g^2 ) \;\;
(g^2)^{\frac{-3 \beta_1}{4 \beta_0^2}} \;\;
e^{-9 S_{PT} / g^2}
\sim
e^{-9 S_{f_B}(g^2) / g^2}
\label{eq:sg}
\ee
where we've now allowed for a $g^2$ dependence in $S_{f_B}$.
Solving eq.(\ref{eq:sg}) for $S_{f_B}(g^2)$, using
the fact that $\epsilon$ is small, we obtain:
\be
S_{f_B}(g^2) \approx S_{PT}
   - \frac{\beta_1}{12 \beta_0^2} g^2 + \frac{\epsilon}{9} g^4.
\label{eq:sss}
\ee
Thus for $\epsilon \approx 0.2$ we get
\be
S_{f_B}(g^2) \approx 1.20 - 0.07 g^2 + 0.02 g^4,
\label{eq:sss1}
\ee
which is not compatible with the values in table \ref{tab:s}.
More significantly, eq.(\ref{eq:sss1}) does not
explain the strong dependence of $S_{f_B}$ on the $g^2$ interval
(see table \ref{tab:s}) or even the fact that $S_i > S_{PT}$.
Thus this analysis suggests that
higher order effects cannot explain the lack of asymptotic scaling of $\fbs$
(ie. $S_{f_B} \ne S_{PT}$) in table \ref{tab:s}.
A similar analysis of the affects of the higher order terms in
$S_i, \;\; i=\{\sigma, M_\rho, f_\pi, 1P-1S\}$ leads to the same conclusion.
The only difference in these cases is that the definition of the $S_i$
in eq.(\ref{eq:a}) includes the $Z^{Ren}_i$ factor,
whereas the definition of $S_{f_B}$ in eq.(\ref{eq:zl}) does not.

The derivation of $S_{f_B}(g^2)$ leading to eq.(\ref{eq:sss}) proves
the statements above which stated that the
functional form chosen in eqs.(\ref{eq:zl},\ref{eq:a}) is quite general
for our purposes. Any $g^2$ dependence in eg. $Z^{Ren}$
can be factored into the definition of $S_i$.
It is also appropriate to note
here that the contribution of $Z^{Ren}$ to $S_{f_B}(g^2)$ is
$\approx \epsilon / 9 \ltap 0.03$ which is much smaller than the
typical statistical errors present in $S_{f_B}(g^2)$ in
table \ref{tab:s}. This justifies ignoring the $g^2$ dependence
of $Z^{Ren}_{f_B}$ in eq.(\ref{eq:ffbb}).

Recently, effects due to higher order terms in lattice perturbation
theory have been addressed \cite{lepage}.
In this work, it has been suggested that $g$, the coupling
constant appearing in the lattice action, is a poor choice
of expansion parameter and the use of
a ``boosted'' coupling, $g_V$, was advocated.
A typical choice of $g_V$ is $\; g^2_V \approx g^2 / u_0^4$,
where $u_0^4$ is the average plaquette.
A straightforward fit of $a_\sigma$
as in eq.(\ref{eq:a}), but with $g$ replaced by $g_V$,
leads to the values of $S^V_\sigma$ in table \ref{tab:s_v},
column 2.

\begin{table}
\begin{center}
\begin{tabular}{ccc}
\hline
\hline
\it{Fitting interval} & $S^V_\sigma$ & $S^V$ from eq.(\ref{eq:s_v}) \\
\hline
\\
$5.7 \le \beta \le 6.0$ & 1.39(2) & 1.07 \\
$6.0 \le \beta \le 6.3$ & 1.26(5) & 1.09 \\
\\
\hline
\hline
\end{tabular}
\caption{ \it{Values for $S^V$ obtained from
(i)  fits to the string tension data in table \ref{tab:a_wil}
to eq.(\ref{eq:a}), but with $g \rightarrow g_V$ (column 2), and,
(ii) a theoretical evaluation using eq.(\ref{eq:s_v}) (column 3).}}
\label{tab:s_v}
\end{center}
\end{table}

A theoretical evaluation based on that leading to
eq.(\ref{eq:sss}), but with $g$ replaced with $g_V$
gives the result:
\bea
S^V(g_V^2) & \approx & S_{PT} - \frac{\beta_1}{12 \beta_0^2} g_V^2
\nonumber \\
           & \approx & 1.20 - 0.07 g_V^2.
\label{eq:s_v}
\eea
$\epsilon$ does not appear in eq.(\ref{eq:s_v}) since
it is zero for the string tension
(ie. $Z^{Ren}_\sigma \equiv 1$).
The values obtained from this formula are shown in the
last column of table \ref{tab:s_v}.
Again, the theoretical predictions do not match the
results from the data. Also it is significant that
the trend of $S^V_\sigma$ with $g^2$ cannot
be reproduced, even with a boosted coupling.

We do not choose to fit the data for $\ia_{M_\rho,f_\pi,1P-1S}$
to the boosted asymptotic scaling prediction since
the larger errors in these cases make the conclusive
interpretation of results difficult.

\subsection{Lattice Artefacts}

The above discussion on higher order terms
is an entirely continuum issue - it does not include any effect
which is purely lattice in origin.
In the following we discuss artefacts of the lattice formulation.
Recalling the discussion in the introduction, we list the
``lattice parameters'' that are involved in these calculations\footnote{
In the past there was concern that $\fbs$ may also be dependant on
the ``smearing'' size of the interpolation operator
used to extract the matrix element \cite{jap}.
However, it is now clear that it is not the case
\cite{sommer,bernard_dallas,fermilab,wup2,ape_6018}.}:
\{ $m_q, L,$ Quenching, $a(g^2)$ \}, where $m_q$ is the (light-) quark
mass and $L$ is the physical extent of the lattice in fermi.
We discuss the effects of each of these parameters in the following
to try to determine the cause of the observed lack of asymptotic
scaling.
 \vskip 10mm
 {\bf ``$m_q$'' effects}
 
 The lattice values of $\fbs$ in this study are all the values
 obtained after an extrapolation in the light quark mass,
 $m_q$, to zero. Thus there is no problem associated with
 the light-quark mass not being adjusted to its physical
 value. One may, however, worry about a systematic error
 due to the extrapolation in $m_q$ \footnote{Normally simulations are
 performed with $m_q \gtap 100$ MeV $>> m_{u,d}$.}. 
 This is unlikely to cause a problem since the dependence
 on $m_q$ of $Z_L$ and indeed $M_\rho, f_\pi,$ and many other quantities
 is mild (see eg. table 3 in \cite{ape_sme0}).
 Presumably also, any systematic effect associated with $m_q$ does
 not depend greatly on $g^2$ over the range studied in this analysis.
 \vskip 10mm
 {\bf Finite Volume effects}
 
\begin{figure}[t]   
 \begin{center}
 \leavevmode
 \epsfysize=540pt
 \epsfbox{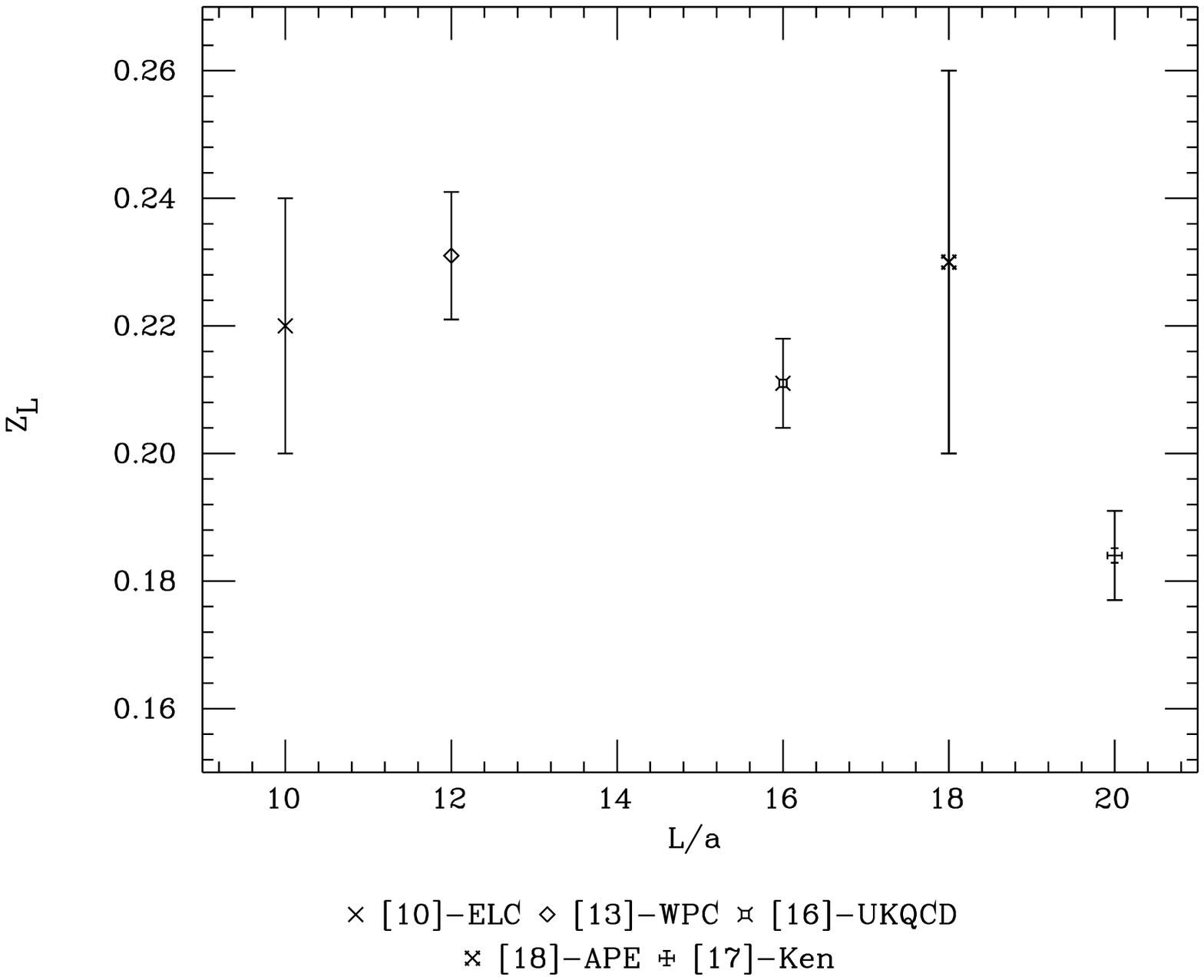}
 \end{center}
 
 \caption[]{\it{Plot of $Z_L$ from various groups at
 $\beta = 6/g^2 = 6.0$ against the spatial dimension in
 lattice units, L/a. The references are as listed in the legend.}}
     \protect\label{fig:zl60}
 \end{figure}

 The effects of finite $L$ on many physical observables has
 been extensively studied \cite{weingarten_new,wup2,wup3}.
 For example, within present statistics, for $L \gtap 1.5 fm$,
 $\fbs$ is not a function of $L$ \cite{wup2}.
 This bound is not entirely satisfied by all the data in this analysis.
 So to study this explicitly, we take the $\zl$ values from table \ref{tab:zl}
 at $\beta = 6.0$ and plot them in fig.(\ref{fig:zl60}) as a function
 of the $L$ used for each simulation.
 There appears if anything to be a {\em decrease} in $Z_L$, and
 therefore correspondingly $\fbs$, as $L$ increases, and this is
 contrary to the observed behaviour \cite{wup2} (see also \cite{BLS2}).
 For this reason, and because generally speaking, different $g^2$ values
 in table \ref{tab:zl} have their lattice sizes chosen such that
 $L \approx const$, we do not believe that finite volume effects
 are to blame for $S_i$ not being equal to its asymptotic freedom
 value, $S_{PT}$.
 
 \vskip 10mm
 {\bf Quenching}
 
 The numerical effects of quenching on lattice calculations are normally
 difficult to uncover, but in the case of $S_i$ this is not the case.
 To see this we note that the quenched approximation is really
 an effective theory of QCD where the sea quark loops are totally
 neglected and the coupling, $g^2$, is adjusted to try to compensate
 for these missing interactions. In an ideal world, this adjustment
 would be perfect, and predictions from
 the correctly-adjusted quenched theory would match those of
 the unquenched theory.
 However, due to the complicated structure of the interactions,
 this is presumably not the case! This means that, for example,
 instead of using the experimental (ie. unquenched) value of
 $\Omega_{M_\rho} = M_\rho = 770$ MeV to set $a_{M_\rho}$,
 we should use the value of $M_\rho$ in a world
 without sea quark loops but where other physical quantities
 (such as $\sigma$ etc.) were as close as possible to their experimental
 values in our unquenched world. That is we should use
 
 \be
 \ia_{M_\rho} = \frac{M_\rho^{Quenched-World}}{M_\rho^L}
 \ee
 cf. eq.(\ref{eq:set}).
 The difference between this prescription and the usual one
 (where instead of $M_\rho^{Quenched-world}$ we have $M_\rho = 770$ MeV)
 is simply a constant pre-factor, independent of $g^2$, and therefore this
 cannot affect $S_{M_\rho}$.
 \vskip 10mm
 {\bf ``O(a)-Effects''}
 
 The final lattice artefact to be discussed is that due to the finiteness
 of the lattice parameter $a$.
 These so-called ``$O(a)$'' effects have long been studied
 (see eg. \cite{sw,symanzik})
 and are known to play an important role in matrix element calculations
 \cite{vlad1,heatlie}.
 They arise because the standard Wilson action replaces continuum
 derivatives by finite
 differences over nearest neighbours and is thus equal to the
 continuum action only up to $O(a)$.
 The clover action is improved to the extent that its predictions are
 correct to $O(a/log(a))$ \cite{heatlie}.
 (Note that the static theory on the lattice is also correct to
 this order \cite{ragazze}.)
 The pure gauge sector of both lattice actions is correct to
 $O(a^2)$, and therefore the string tension, which is a pure gauge
 quantity, is correct to this order.
 
 Summarizing this discussion, we have for the Wilson case:
 
 \be
 \Omega_i^L = \Omega_i^{PL}+ O(a), \\
 \label{eq:ordera}
 \ee
 and for the clover case:
 \be
 \Omega_i^L = \Omega_i^{PL}+ O(a/log(a)), \\
 \label{eq:orderaloga}
 \ee
 where $i = f_B,M_\rho,f_\pi,1P-1S$,
 and for the pure gauge case:
 \be
 \Omega_{\sigma}^{L} = \Omega_{\sigma}^{PL} + O(a^2).
 \label{eq:ordera2}
 \ee
 The $\Omega_i^{PL}$ are the results that would be obtained
 with a perfect lattice action, ie. one correct to all orders in $a$.
 Using eqs.(\ref{eq:ordera} \& \ref{eq:ordera2}) together with
 eqs.(\ref{eq:set} \& \ref{eq:a_PT}), we obtain, in the Wilson case for
 $i = {M_\rho,f_\pi,1P-1S}$,
 \be
 \ia_i(g^2) = \Lambda_i \;\; f_{PT}^{-1}(g^2) \;\;
              ( 1 - X_i \frac{f_{PT}(g^2)}{f_{PT}(1)} ),
 \label{eq:a_fit1}
 \ee
 and, for $i = \sigma$,
 
 \be
 \ia_\sigma(g^2) = \Lambda_\sigma \;\; f_{PT}^{-1}(g^2) \;\;
 ( 1 - X_\sigma ( \frac{f_{PT}(g^2)}{f_{PT}(1)} )^2 ),
 \label{eq:a_fit2}
 \ee
 where $X_i$ is the relative
 strength of the $O(a)$ (or $O(a^2)$) correction at
 $g^2 = 1$, ie. $\beta = 6/g^2 = 6.0$.
 Here we have replaced $a_{PL}$ with $\Lambda^{-1} f_{PT}$
 (see eq.(\ref{eq:a_PT})) since $a_{PL} = \Lambda^{-1} f_{PT}$
 up to exponentially suppressed non-perturbative pieces.
 
 Thus we see straightaway that in the limit where the effects
 of quenching are unimportant (and therefore $\Lambda_i = \Lambda_j$)
 $\ia_i(g^2) - \ia_j(g^2) = const.\;$ for fermionic Wilson quantities.
 In the future, with better statistics, this can be checked.
 
 We are now in a position to fit the data in table \ref{tab:a_wil}
 to the appropriate eqs.(\ref{eq:a_fit1} \& \ref{eq:a_fit2}).
 In this fit there are two free parameters: $X_i$ and the coefficient,
 $\Lambda_i$.
 
\begin{figure}[t]   
    \begin{center}
    \leavevmode
    \epsfysize=540pt
    \epsfbox{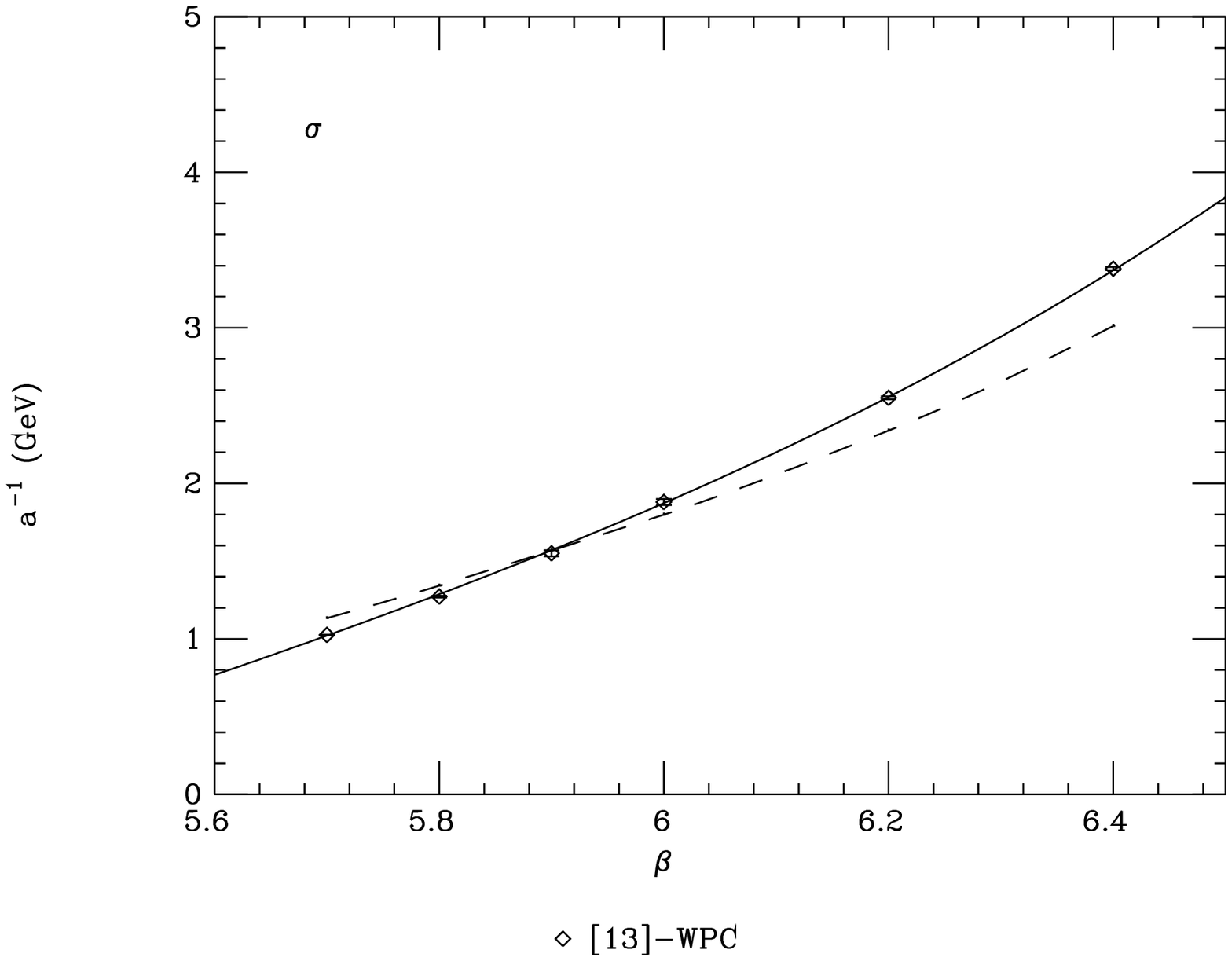}
    \end{center}
    
    \caption[]{\it{Plot of $\ia$ from the string tension against $\beta = 6/g^2$.
    The solid curve is the fit to eq.(\ref{eq:a_fit2})
    (ie. the 2-loop asymptotic scaling formulae with an $O(a^2)$ term).
    The dashed curve is the fit to eq.(\ref{eq:a_PT}) with
    $g^2$ replaced by $g_V^2$
    (ie. the ``boosted'' 2-loop asymptotic scaling formulae).
    The references are as appears in the legend.}}
    \protect\label{fig:iaa_a}
    \end{figure}

 Taking all the $\ia_\sigma(g^2)$ values in table \ref{tab:a_wil}
 and fitting them to eq.(\ref{eq:a_fit2}) we obtain
 $X_\sigma = 0.197(2)$ with a $\chi^2/dof$ of $11/4$
 which is shown in table \ref{tab:X}.
 As a check we also fit $\ia_\sigma(g^2)$ to eq.(\ref{eq:a_fit1})
 obtaining, as expected, a poorer $\chi^2/dof$ of $26/4$
 with $X_\sigma = 0.343(4)$.
 This suggests that we are in fact isolating the $O(a^2)$ corrections
 in $\sqrt\sigma$, and that furthermore, the largish value
 of $\chi^2/dof$ in the fit to eq.(\ref{eq:a_fit2}) may signal the
 statistical presence of even higher order terms (ie. of $O(a^n), n>2$).
 The fit of $\ia_\sigma(g^2)$ to eq.(\ref{eq:a_fit2}) is shown
 in fig.(\ref{fig:iaa_a}) as a solid line.
 In this figure we have also shown, as a dashed curve,
 a fit to the 2-loop asymptotic scaling formulae with
 the boosted coupling $g_V$ as discussed in sec.4.1
 (ie. eq.(\ref{eq:a_PT}) with $g^2 \rightarrow g_V^2 = g^2 / u_0^4$,
 where $u_0^4$ is the average plaquette).
 As can be seen from fig.(\ref{fig:iaa_a}),
 the quality of this boosted asymptotic scaling fit is
 poor; in fact the $\chi^2/dof$ for this fit is around $10^3$.
 On the other hand, the fit to eq.(\ref{eq:a_fit2})
 (ie. the asymptotic scaling formulae with an $O(a^2)$ term)
 is very acceptable.
 This suggests strongly that the observed lack of
 asymptotic scaling in the string tension data
 can be simply explained by $O(a^2)$-effects.
 
\begin{figure}[t]   
    \begin{center}
    \leavevmode
    \epsfysize=540pt
    \epsfbox{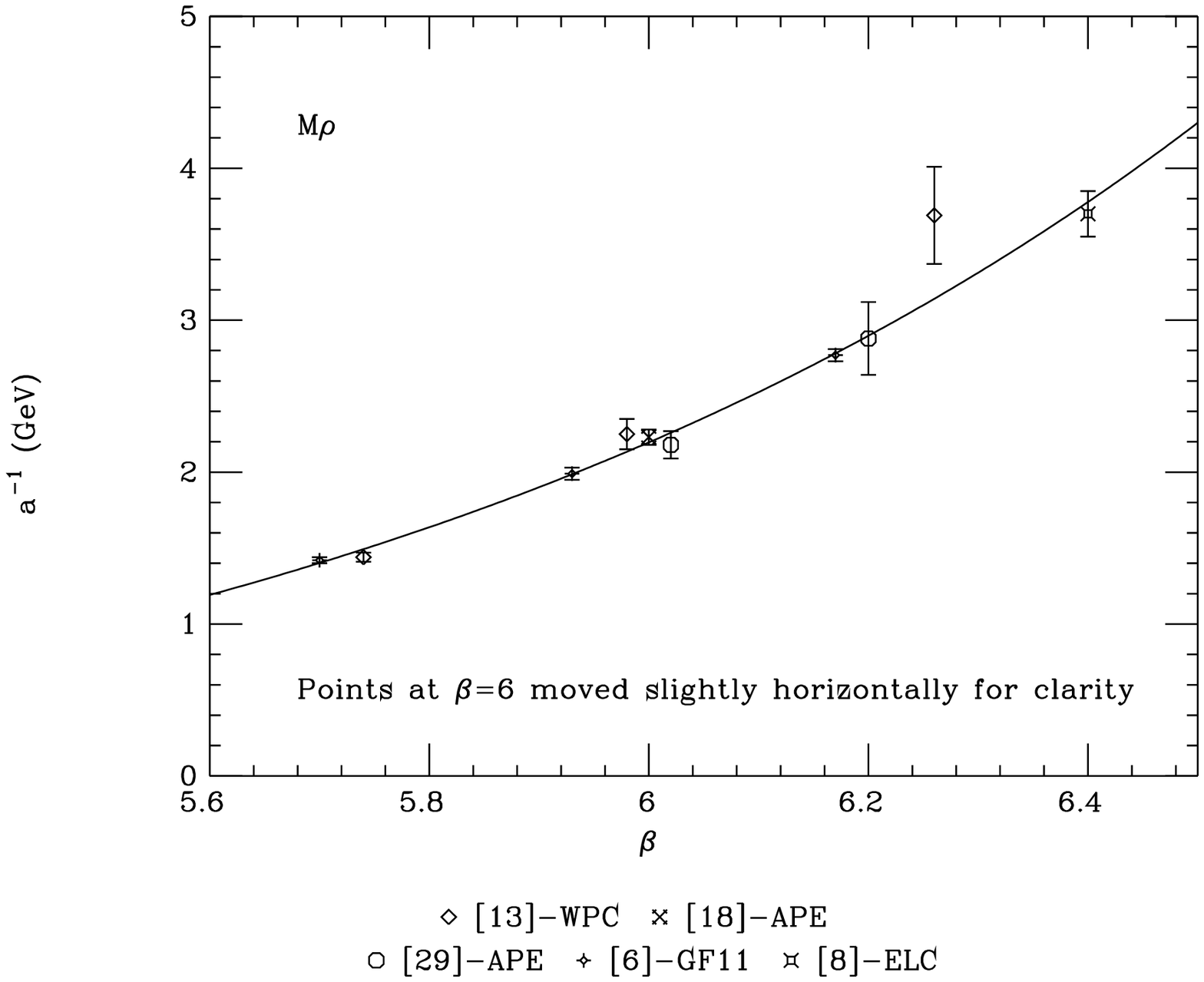}
    \end{center}
    
    \caption[]{\it{Plot of $\ia$ from $M_\rho$ against $\beta = 6/g^2$.
    The solid curve is the fit to eq.(\ref{eq:a_fit1})
    (ie. the 2-loop asymptotic scaling formulae with an $O(a)$ term).
    The references are as appears in the legend.
    All data is from the Wilson action.}}
    \protect\label{fig:iab_a}
    \end{figure}

\begin{figure}[t]   
    \begin{center}
    \leavevmode
    \epsfysize=540pt
    \epsfbox{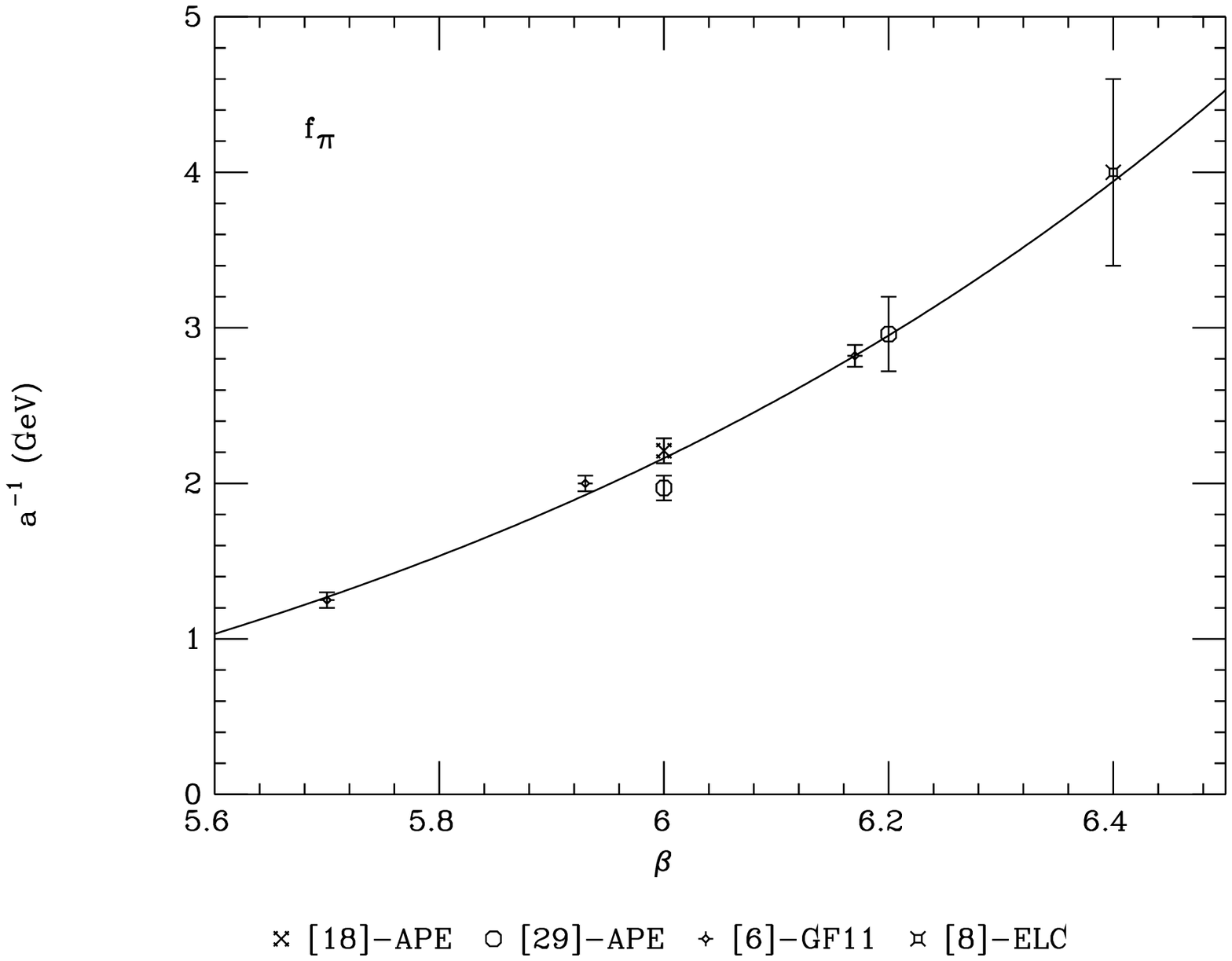}
    \end{center}
    
    \caption[]{\it{Plot of $\ia$ from $f_\pi$ against $\beta = 6/g^2$.
    The solid curve is the fit to eq.(\ref{eq:a_fit1})
    (ie. the 2-loop asymptotic scaling formulae with an $O(a)$ term).
    The references are as appears in the legend.
    All data is from the Wilson action.}}
    \protect\label{fig:iac_a}
    \end{figure}

\begin{figure}[t]   
    \begin{center}
    \leavevmode
    \epsfysize=540pt
    \epsfbox{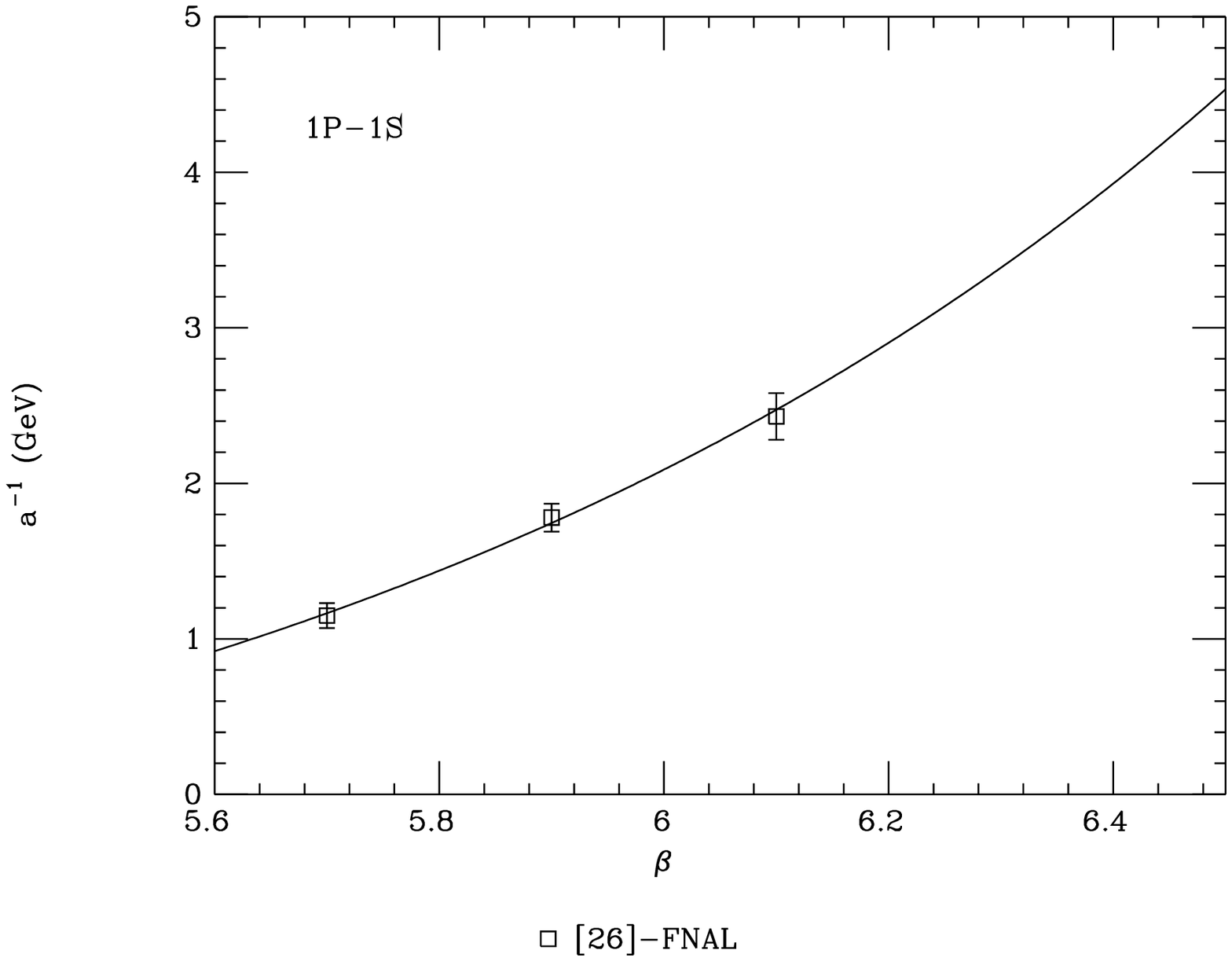}
    \end{center}
    
    \caption[]{\it{Plot of $\ia$ from the $1P-1S$ splitting in charmonium
    against $\beta = 6/g^2$.
    The solid curve is the fit to eq.(\ref{eq:a_fit1})
    (ie. the 2-loop asymptotic scaling formulae with an $O(a)$ term).
    The references are as appears in the legend.
    All data is from the Wilson action.}}
    \protect\label{fig:iad_a}
    \end{figure}

 In the case of $\ia_{M_\rho,f_\pi,1P-1S}$ for the Wilson
 data, a fit to eq.(\ref{eq:a_fit1})
 gives the values for $\Lambda_i$ and $X_i$ shown in table \ref{tab:X}.
 These fits are shown in figs.(\ref{fig:iab_a}-\ref{fig:iad_a}).
 Again the $\chi^2/dof$ are very acceptable.
  
 \begin{table}
 \begin{center}
 \begin{tabular}{ccccc}
 \hline
 \hline
 $i$              & $\sigma$ & $M_\rho$  & $f_\pi$  & 1P-1S \\
 fit to eq.       & \ref{eq:a_fit2} &
                    \ref{eq:a_fit1} & \ref{eq:a_fit1} & \ref{eq:a_fit1} \\
 \hline
 $X_i$            & 0.197(2) & 0.21(2)   & 0.31(5)  & 0.35(11)\\
 $\Lambda_i$ (MeV)& 1.780(5) & 2.12(6)   & 2.4(1)   & 2.5(3)  \\
 $\chi^2_i/dof$   & 11/4     & 8/8       & 8/5      & 0.3/1   \\
 \hline
 \hline
 \end{tabular}
 \caption{ \it{Values for $X_i, \Lambda_i$ and $\chi^2/dof$ obtained from
 fits of the Wilson data in table \ref{tab:a_wil}
 to eqs.(\ref{eq:a_fit1},\ref{eq:a_fit2}) as indicated.}}
 \label{tab:X}
 \end{center}
 \end{table}
 
 Due to the quality of the fits we conclude that the most satisfactory
 explanation of the observed lack of {\em asymptotic} scaling
 (ie. $S_{\sigma,M_\rho,f_\pi,1P-1S} \ne S_{PT}$) is $O(a)$ effects.
 This is the only explanation out of those discussed here
 which seems consistent with the data.
 
 The results for the $\Lambda$ values in table \ref{tab:X}
 indicate that $\Lambda_\sigma$ is significantly lower than
 the Wilson values for $\Lambda_{M_\rho,f_\pi,1P-1S}$,
 and that therefore results
 using $\ia_\sigma$ will differ statistically in the continuum limit
 from those using $\ia_{M_\rho,f_\pi,1P-1S}$. However, $\sigma$
 itself is in fact a poorly determined quantity since it relies
 on model calculations, and furthermore, from above arguments,
 the low value of $\Lambda_\sigma$ may signal the effect of quenching.
 
 For the clover case, we do not attempt to extract the
 coefficient of the $a / log(a)$ term due to the fact that there are
 data at only two values of $g^2$ available. We await further data before
 attempting this analysis.
 Also, we choose not to perform a combined least squares fit of
 $\ia_{M_\rho,f_\pi,1P-1S}$ to eq.(\ref{eq:a_fit1}) together with
 $\ia_{\sigma}$ to eq.(\ref{eq:a_fit2})
 with a {\em single} $\Lambda$, since quenching implies that a single
 $\Lambda$ is inappropriate.
 
\begin{figure}[t]   
    \begin{center}
    \leavevmode
    \epsfysize=540pt
    \epsfbox{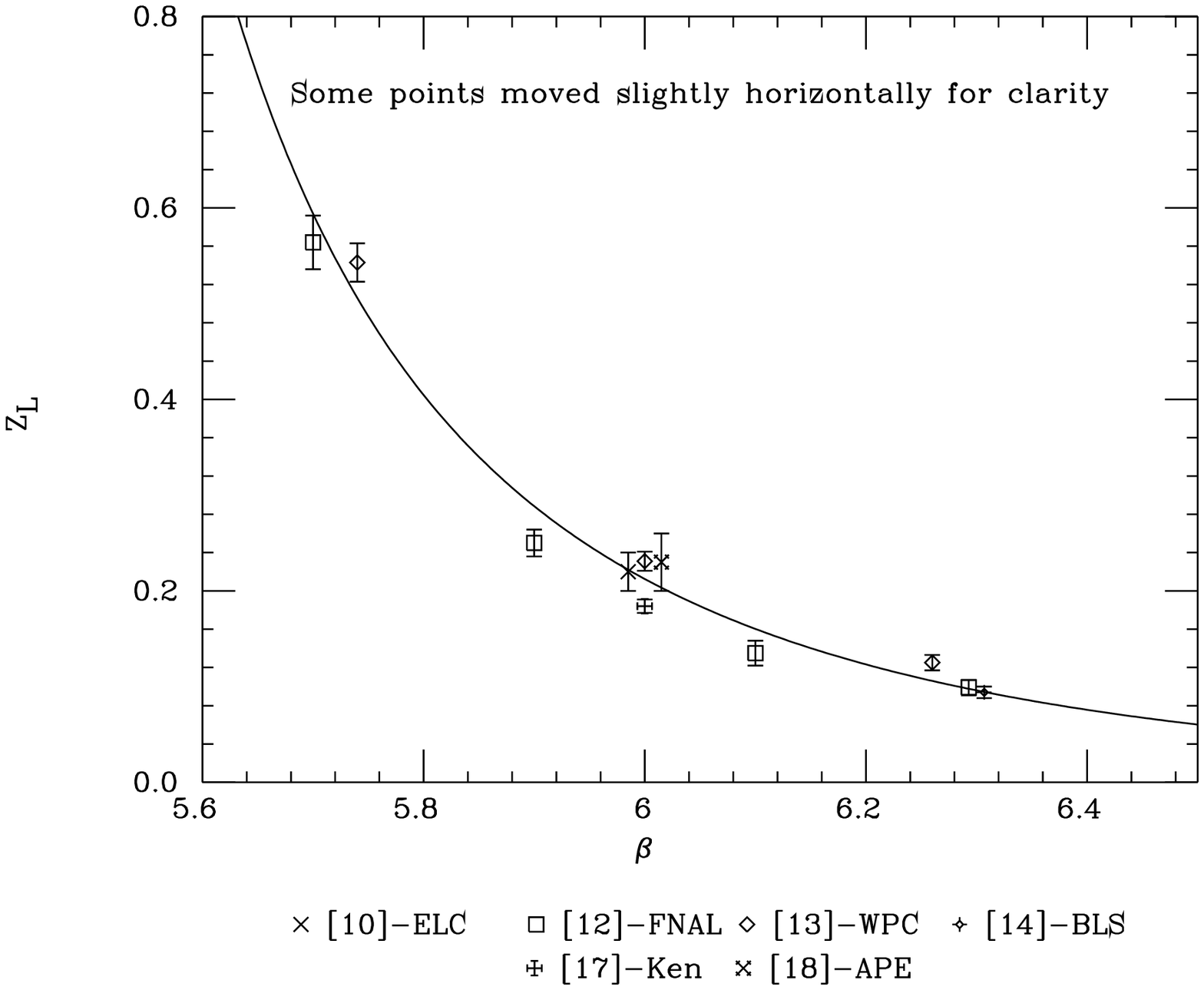}
    \end{center}
    
    \caption[]{\it{Plot of $Z_L$ against $\beta = 6/g^2$.
    The solid curve is the fit to eq.(\ref{eq:a_zl})
    (ie. the 2-loop asymptotic scaling formulae with an $O(a)$ term).
    The references are as appears in the legend.
    All data is from the Wilson action.}}
    \protect\label{fig:zl_ordera}
    \end{figure}

 We can continue the analysis of the $O(a)$ effects by
 fitting $Z_L$ as follows:
 
 \be
 \frac{1}{Z_L^{2/3}} = \lambda_{Z_L} \;\; f_{PT}^{-1}(g^2) \;\;
                    ( 1 - X_{Z_L} \frac{f_{PT}(g^2)}{f_{PT}(1)} ).
 \label{eq:a_zl}
 \ee
 We obtain the values of $X_{Z_L}$ and $\lambda_{Z_L}$ in
 table \ref{tab:X_zl}. (Again we have fitted only the Wilson data.)
 Note that this functional form was chosen
 to mirror the fits of $\ia_{\sigma,M_\rho,f_\pi,1P-1S}$ to
 eqs.(\ref{eq:a_fit1} \& \ref{eq:a_fit2}). However, of course,
 in this case the physical value corresponding to $Z_L$ (ie. $\fb$)
 is not known, and there is nothing to set the scale;
 therefore $\lambda_{Z_L}$ is dimensionless.
 In fig.(\ref{fig:zl_ordera}) we plot the $Z_L$ data against
 $\beta$ (for the Wilson action) together with the fit to
 eq.(\ref{eq:a_zl}) shown as a solid line.
 Note from this plot and from the relatively poor $\chi^2$ in
 table \ref{tab:X_zl} that there appears to be some systematic
 effects remaining in the data. This is not entirely surprising
 due to the known difficulty of extracting $Z_L$.
 
 \begin{table}
 \begin{center}
 \begin{tabular}{cc}
 \hline
 \hline
 fit to eq.       & \ref{eq:a_zl}  \\
 \hline
 $X_{Z_L}$        & 0.42(3) \\
 $\lambda_{Z_L}$  & $3.7(1) \times 10^{-3}$  \\
 $\chi^2_i/dof$   & 39/9    \\
 \hline
 \hline
 \end{tabular}
 \caption{ \it{Values for $X_{Z_L}, \lambda_{Z_L}$ and $\chi^2/dof$
 obtained from a fit of the $Z_L$ values in table \ref{tab:zl}
 to eq.(\ref{eq:a_zl}).}}
 \label{tab:X_zl}
 \end{center}
 \end{table}

 
 We are now in a position to obtain a continuum value of $\fbs$.
 Using
 eqs.(\ref{eq:fbs_define},\ref{eq:a_fit1},\ref{eq:a_fit2},\ref{eq:a_zl}),
 in the limit $a \rightarrow 0$
 
 \be
 \fbs(a=0) \;\; = \;\; \sqrt{\frac{2}{M_B}} \;\; Z^{Ren}_{f_B}(g^2=0) \;\;
 (\frac{\Lambda_i}{\lambda_{Z_L}})^{3/2}
 \label{eq:fbs_cont}
 \ee
 where in this limit,
 $Z^{Ren}_{f_B}(g^2=0) = 1 - \frac{3S_{PT}}{2\pi^2} = 0.82$
 We use the average of the $\Lambda_{M_\rho,f_\pi,1P-1S}$
 from table \ref{tab:X} ie. $\Lambda = 2.2(2)$ MeV (where the
 error is statistical plus systematic combined in quadrature)
 and obtain,
 \be
 \fbs(a=0) \;\; = \;\; 230(35) {\rm MeV}.
 \ee
 We take this as our best estimate of the continuum value of $\fbs$
 from the lattice. 
 Note that this is roughly equivalent to the values
 obtained from simulations at finite $a$ values for $\ia \gtap 2$ GeV
 (see fig.(\ref{fig:bernard}), note though that this figure uses the
 scale from the string tension).
 Had we instead used $\Lambda_\sigma = 1.780(5)$,
 we would obtain $\fbs(a=0) = 170(14)$ MeV. We do not prefer to
 choose this value since it appears that the $\ia_\sigma$ values
 are contaminated by either quenched effects, or model dependences.
 
 Since the renormalization constant $Z^{Ren}_{f_B}$ between the
 lattice and continuum effective theories is evaluated
 at $g^2 = 0$, it is exactly determined.
 This is in contrast with the uncertainties in 
 $Z^{Ren}_{f_B}$ at finite $g^2$ which plague other approaches
 due to the uncalculated terms of order $g^4$ and higher.
    
\section{Conclusion}

In this paper we have presented a method to obtain the continuum
value of $\fbs$ from lattice data. This approach separately isolates the
systematic errors coming from the dimensionless lattice
quantity corresponding to $\fbs$ (ie. $Z_L$) and the lattice quantity
used to determine the scale. These systematic errors and the lack of
asymptotic scaling can be parameterised
in terms of finite lattice spacing effects.
Assuming this explanation a value of $\fbs = 230(35)$ MeV
in the continuum limit has been obtained.
In this method there is no error in the renormalization constant
connecting the lattice and continuum effective theories.

\section{Acknowledgements}

It is a pleasure to thank G. Martinelli and C. Sachrajda
for guidance and support over the last years.
The author also wishes to acknowledge many useful discussions
with colleagues in the APE collaboration,
especially M. Crisafulli, V. Lubicz, F. Rapuano and A. Vladikas,
and for allowing the use of some unpublished APE data in table 4.
We thank the INFN for the financial support which has been forthcoming
and the European Union for support under the Human Capital and Mobility
Program, contract ERBCHBICT941462.


\end{document}